\documentclass[preprint]{elsarticle}
\usepackage{appendix}
\usepackage{amssymb}
\usepackage{amsmath}
\usepackage[final]{epsfig}
\usepackage{color}
\usepackage{graphicx}
\usepackage{lineno}
\usepackage{bm}
\usepackage{comment}
\usepackage{braket}
\usepackage{dsfont}
\usepackage{placeins} 
\usepackage{slashed}
\newcommand{\beq}{\begin{equation}}
\newcommand{\eeq}{\end{equation}}
\newcommand{\bdi}{\begin{displaymath}}
\newcommand{\edi}{\end{displaymath}}
\newcommand{\no}{\nonumber}
\newcommand{\bea}{\begin{eqnarray}}
\newcommand{\eea}{\end{eqnarray}}

\newcommand{\ov}{\overline}

\newcommand{\de}{\partial}

\newcommand{\comm}[1]{}

\begin{document}

\begin{frontmatter}

\title{Time resolution and efficiency of SPADs and SiPMs for photons and charged particles}

\author[CERN]{W. Riegler}
\author[Oxford]{P. Windischhofer}

\address[CERN]{CERN}
\address[Oxford]{University of Oxford}

\begin{abstract}
  
We give an analytic treatment of the time resolution and efficiency of Single Photon Avalanche Diodes (SPADs) and Silicon Photomultipliers (SiPMs). We provide closed-form expressions for structures with uniform electric fields and efficient numerical prescriptions for arbitrary electric field configurations. We discuss the sensor performance for single photon detection and also for charged particle detection. \\

\end{abstract}

\end{frontmatter}

\newpage

\section{Introduction}

Single Photon Avalanche Diodes (SPADs) have been used for photon detection and photon counting since many decades.  These semiconductor devices contain a highly doped p-n junction of $0.5{-}2\,\mu$m thickness, the so called gain region, that is biased above the breakdown voltage. This means that a single primary electron or hole entering this region can produce a diverging avalanche through impact ionization and therefore lead to a detectable signal. The growth of the diverging avalanche is quenched by the breakdown of the electric field, which leads to a digital type signal with an amplitude independent of the number of primary charges. The restoration of the electric field is governed by a quench resistor or quench circuit external to the device. \\
\begin{figure}[h]
\begin{center}
     \epsfig{file=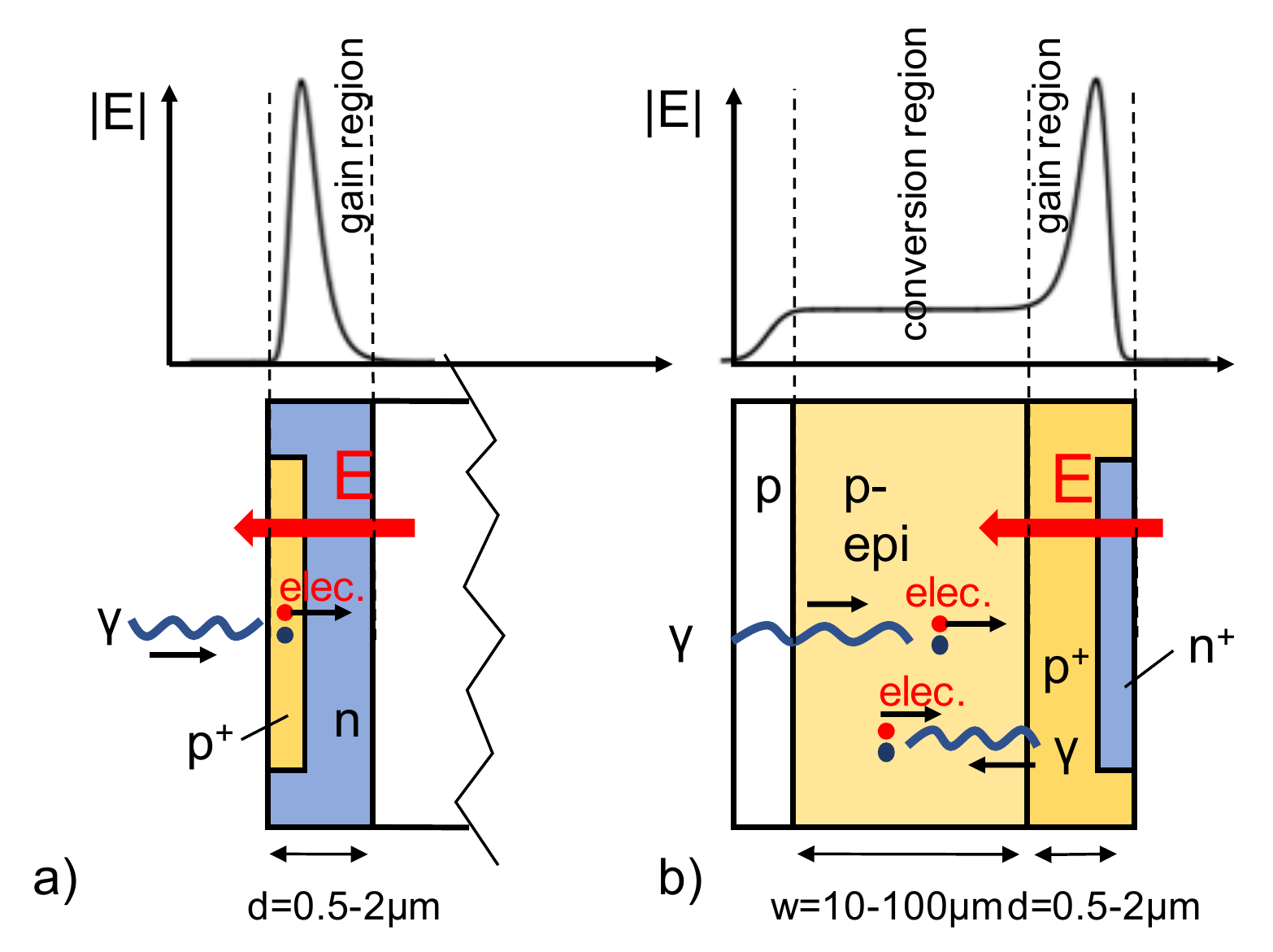, height=7cm}
\caption{a) p-in-n SPAD with a gain layer only. b) n-in-p SPAD with a gain layer and a conversion layer.}
\label{real_spad_schemes}
\end{center}
\end{figure}
\\
Two basic SPAD geometries are shown in Fig.~\ref{real_spad_schemes}. 
For the p-in-n SPAD in Fig.~\ref{real_spad_schemes}a the high field region is placed close to the surface of the silicon. Photons with a wavelength of $<500$\,nm have an absorption length of less than $1\,\mu$m, so they will be efficiently absorbed in the gain region and trigger the breakdown.
 \\
 For photons with longer wavelengths and therefore larger absorption length, the geometry of Fig.~\ref{real_spad_schemes}b with a so called 'conversion region' will be more efficient. This region consists of a depleted layer of silicon with thickness ranging from 10-100\,$\mu$m that is adjacent to a gain region. The electrons created in the conversion region will drift to the gain region where they provoke the breakdown. The sensor can be illuminated from the front and the back side. These SPADs are nowadays heavily used for LIDAR \cite{LIDAR} applications with wavelengths in the near-infrared above 750\,nm.
\\ \\
While originally being a single channel device, the advances in silicon industry allowed the arrangement of many of these diodes in chessboard structures. Depending on the readout circuitry these are called SPAD image sensors or Silicon Photomultipliers (SiPMs). In SPAD image sensors, the SPAD pixels are read out individually in order to form an image. In SiPMs the individual pixels are connected in parallel through series resistors into a single channel. The amplitude of the output signal will then be proportional to the total number of fired pixels and therefore proportional to the number photons that have hit the pixel matrix, which represents the function of a traditional photomultiplier. 
There are many technological challenges related to the implementation of such pixel structures, specifically the elimination of crosstalk between the channels and the minimization of the dark count rate. 
\\ \\
{\bf Scope and outline} \\
In this report, we discuss the time resolution and efficiency of SPADs and SiPMs of the two types shown in Fig.~\ref{real_spad_schemes} and evaluate their performance for the detection of photons as well as charged particles. Our results are derived from a series of fundamental equations that describe the movement and the interactions of the participating charge carriers. \\
We provide analytic expressions for the simplified structures in Fig.~\ref{spad_scheme} with constant electric fields. 
Fig.~\ref{spad_scheme}a represents the situation where a photon interacts directly inside a  gain layer of constant electric field $E_1$ that is above the breakdown value. 
Fig.~\ref{spad_scheme}b shows a SPAD where we represent the conversion region by a silicon layer of thickness $w$ with constant electric field $E_0$ below the breakdown field, which is adjacent to a gain layer with constant field $E_1$.
\begin{figure}[h]
\begin{center}
     \epsfig{file=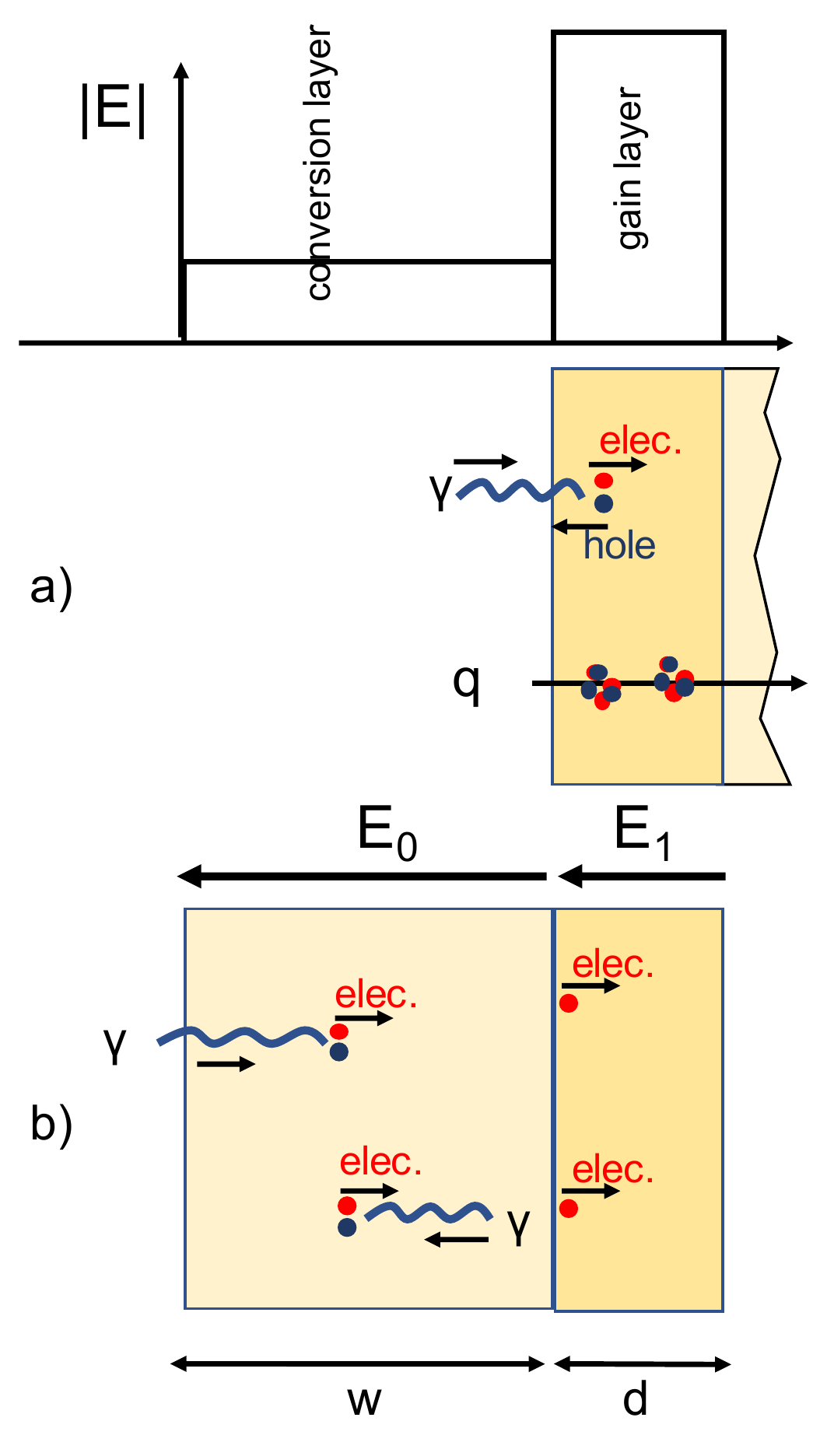, width=5cm}
\caption{
  Two simplified SPAD structures with constant electric fields discussed in this report: a) A photon interacts directly in the gain layer and produces an e-h pair that provokes breakdown. b) A photon produces an e-h pair in the conversion layer and the electron moves to the gain layer where it provokes breakdown.  A thin gain layer of 0.5-2\,$\mu$m is also highly efficient for charged particle detection.}
\label{spad_scheme}
\end{center}
\end{figure}
 \\ 
This report is structured as follows: Section 2 discusses the absorption of photons in silicon, and the contribution of the conversion layer to the time resolution. 
Section 3 then discusses the mechanisms of avalanche breakdown and the average growth of the avalanche. Section 4 describes the contribution to the efficiency resulting from the avalanche formation in the gain layer. Section 5 discusses avalanche fluctuations and their contribution to the time resolution. This discussion is based to a large degree on the companion paper \cite{philipp_paper}, which develops the statistics of electron-hole avalanches in great detail. Section 6 then discusses the performance of SPADs for the detection of charged particles.
We finally drop the assumption of constant electric field in Section 7 and return to the realistic field configurations of Fig.~1. We give efficient numerical prescriptions that extend the analytic results obtained previously.  \\
Details of all calculations are given in several appendices. Although we focus specifically on devices based on silicon, our results are expected to also cover the basic geometries for different types of semiconductors. 


\section{Conversion layer \label{conversion_layer}} 

\begin{figure}[h]
\begin{center}
     \epsfig{file=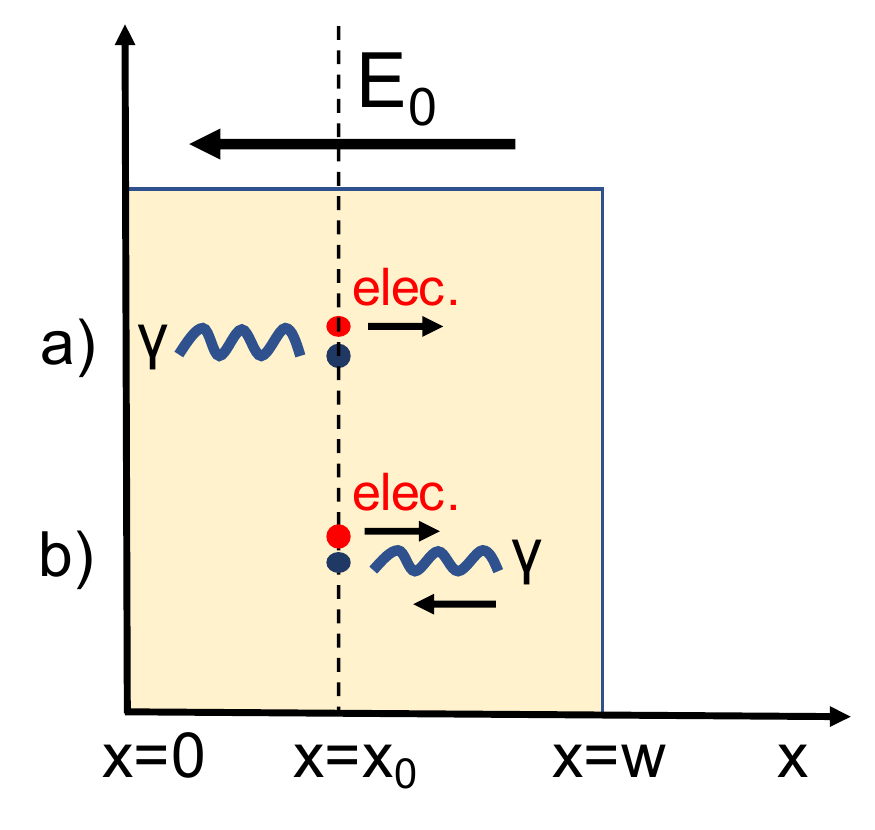, width=5cm}
     \caption{The conversion layer of a SPAD or SiPM. A photon is absorbed at position $x_0$ producing an e-h pair. The electron will drift to the gain layer at $x=w$.   }
\label{drift_layer1}
\end{center}
\end{figure}

\begin{figure}[h]
\begin{center}
      a) \epsfig{file=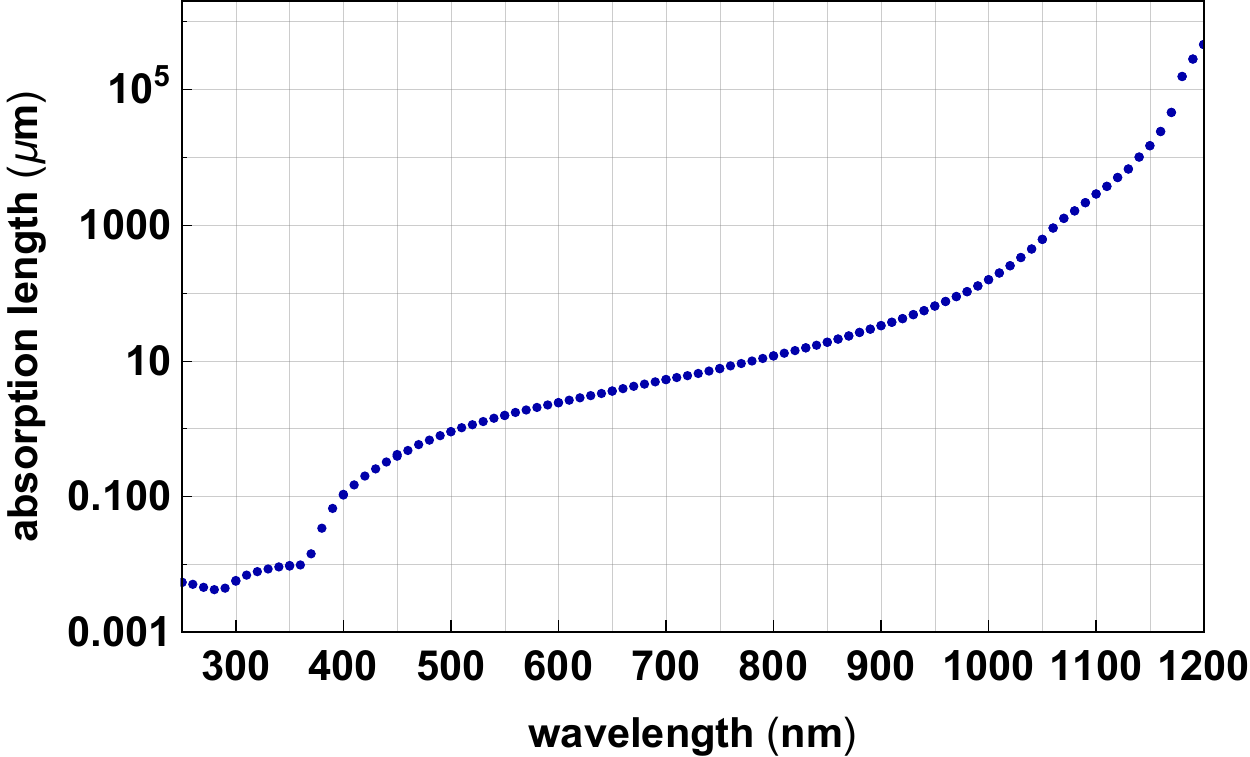, width=6cm}
      b)  \epsfig{file=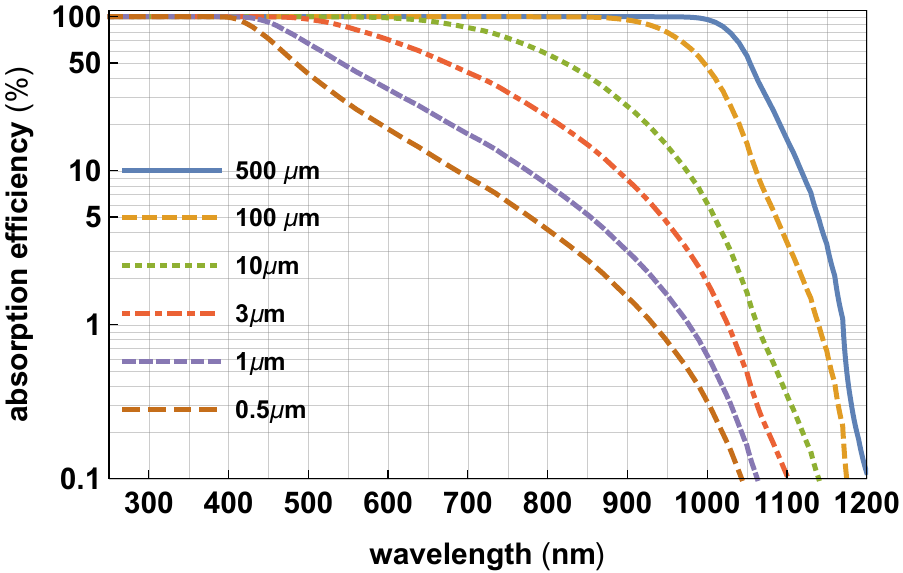, width=6cm}
\caption{ Absorption length $l_a$ for photons of different wavelengths in silicon \cite{green1, green2}. b) Photon absorption efficiency for different values of silicon thickness.}
\label{absorption_length}
\end{center}
\end{figure}

We assume a layer of silicon of thickness $w$ extending from $x=0$ to $x=w$ as shown in Fig.~\ref{drift_layer1}.  The probability for a photon to be absorbed between position $x_0$ and $x_0+dx_0$ is given by $P(x_0)dx_0 = 1/l_a e^{-x_0/l_a} dx_0$, where $l_a$ is the photon absorption length from Fig.~\ref{absorption_length}a. The efficiency, i.e.~the probability for a photon to convert in the layer, is then given by $p=1-e^{-w/l_a}$ and the numbers are shown in Fig.~\ref{absorption_length}b. Photons of wavelength $<500$\,nm are efficiently absorbed in $<1\,\mu$m of silicon while infrared photons of ${>}750$\,nm need several tens of $\mu$m of silicon to be absorbed efficiently. 
\\
First we investigate the case where the photon is arriving from the 'left' side as shown in Fig.~\ref{drift_layer1}a. Normalising the conversion probability to the efficiency, the probability for a photon to be absorbed between position $x_0$ and $x_0+dx_0$ is 
\beq \label{conversion_probability} 
     P(x_0)dx_0 = \frac{1}{1-e^{-w/l_a}}\frac{1}{l_a} e^{-x_0/l_a} \Theta(w-x_0)dx_0
\eeq
The electron is then moving to the edge of the silicon layer at $x=w$ with a velocity $v_e$, where it arrives at time $t=(w-x_0)/v_e$. The velocity of electrons and holes in silicon is shown in Fig.~\ref{drift_velocity}a and the parametrization is given in Appendix A.
\begin{figure}[h]
\begin{center}
a)
     \epsfig{file=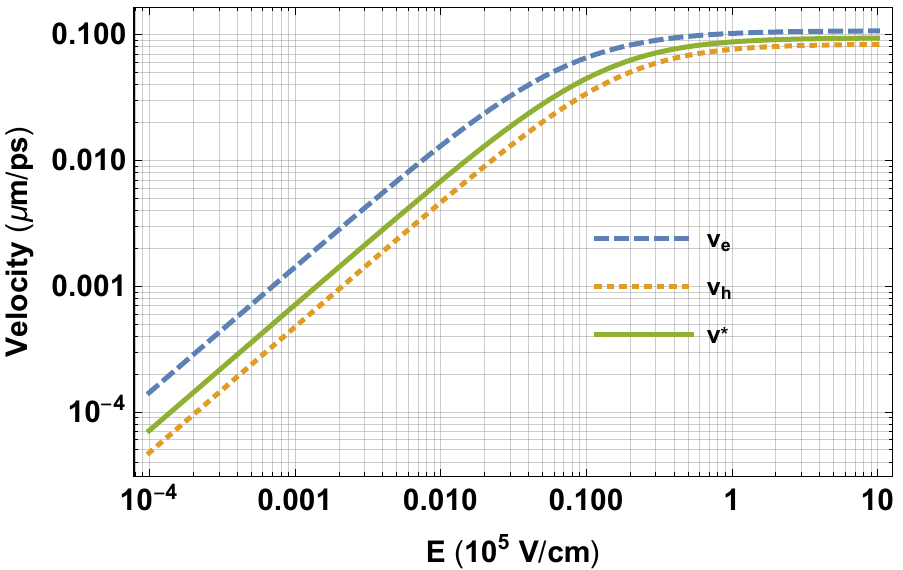, width=6cm}
 b)
      \epsfig{file=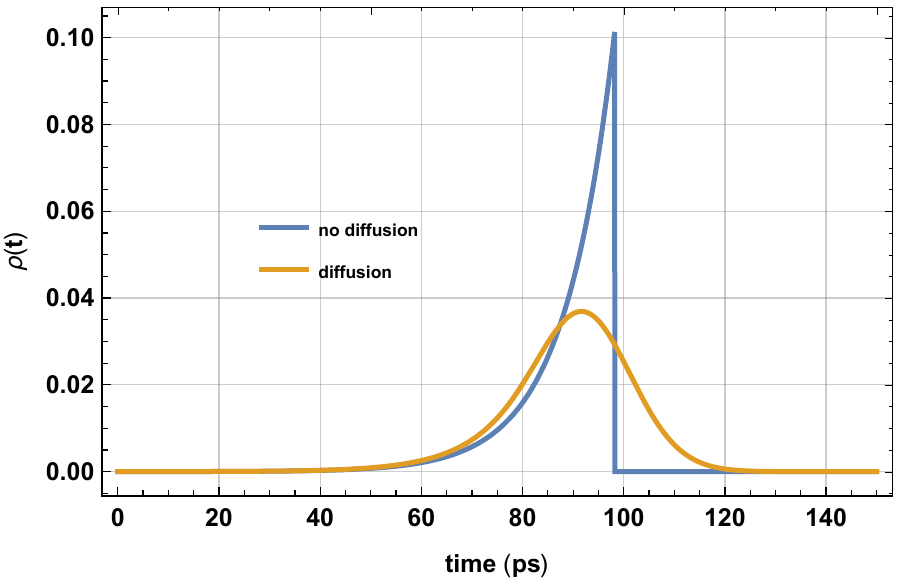, width=6cm}
\caption{a) Drift velocity of electrons ($v_e$) and holes ($v_h$) as a function of electric field in silicon. The velocity $v^*=2v_ev_h/(v_e+v_h)$ that is relevant for the avalanche growth in the gain layer is shown as well. b) Probability for the electron to arrive at $x=w$ between times $t$ and $t+dt$ for $w=10\,\mu$m, for a photon with $\l_a=1\,\mu$m entering the layer from the left side. The 'no diffusion' curve refers Eq. \ref{rho1} and the 'diffusion' curve refers to Eq. \ref{rho2}.}
\label{drift_velocity}
\end{center}
\end{figure}
The arrival time distribution of the electron at $x=w$ is therefore
\beq \label{rho1}
      \rho_1(t) = \int_0^w P(x_0) \delta[ t - (w-x_0)/v_e ] dx_0 =  \frac{w}{l_a (e^{w/l_a}-1)}\frac{1}{T} e^{\frac{t w}{l_a T}} \,\Theta(T-t)
\eeq
where we have expressed the velocity $v_e$ by the maximum drift time  $T=w/v_e$ of the electrons inside the conversion layer. An example is shown in  Fig.~\ref{drift_velocity}b. The variance of the arrival time is then
\beq \label{drift_layer_position}
  \sigma_t ^2= \int_0^T t^2\,P_2(t) dt -  \left( \int_0^T t\,P_2(t) dt \right)^2 = T^2 \left(\frac{l_a^2}{w^2} - \frac{1}{4  \sinh[w/2\l_a]^2}\right)
\eeq
Including the effect of diffusion we use the fact that an electron deposited at position $x_0$ at $t=0$ and moving with an average velocity of $v_e$ will be found at position $x$ after a time $t$ with a probability of
\beq \label{gaussian}
    p(x, x_0, t) dx = \frac{1}{\sqrt{2 \pi} \sqrt{2Dt}} \exp \left[ -\frac{(x-(x_0+v_et))^2}{2(2Dt)} \right] dx
\eeq
The standard deviation of the distribution is given by $\sigma(t) = \sqrt{2Dt}$. The probability for an electron to arrive at $x=w$ between time $t$ and $t+dt$ is then $p(w, x_0, t) v_e dt$. Since the probability of a photon absorption at $x_0$ is given by $P(x_0)$  from Eq.~\ref{conversion_probability} we find the arrival time distribution $P_3(t)$ as 
\beq
       \rho_2 (t) = \int_0^w P(x_0) p(w, x_0, t) v_e dx_0 = 
\eeq
\beq \label{rho2}
      \frac{v_e}{2l_a} \frac{ 1 }{e^{w/l_a}-1 }  e^{(D+l_a v_e)t/l_a^2} 
      \left(
                     \mbox{Erf} 
                 \left[    
                    \left(    \frac{\sqrt{Dt}}{l_a}+\frac{v_e}{2} \sqrt{\frac{t}{D}}    \right)
               \right] -
                 \mbox{Erf} 
                 \left[    
                  \frac{\sqrt{Dt}}{l_a} + \frac{v_e}{2} \sqrt{\frac{t}{D}}     -  \frac{w}{2 \sqrt{Dt}}
               \right]               
    \right)
\eeq
with Erf$(z) = 2/\sqrt{\pi} \int_0^z e^{-t^2}dt$.  The variance evaluates to
\beq
       \sigma_t^2 =  T^2 \left(\frac{l_a ^2}{w^2} - \frac{1}{4  \sinh[w/2l_a ]^2}\right) +T\frac{2D}{v_e^2}\left( \frac{1}{1-e^{-w/l_a }} -\frac{l_a }{w}\right) + \frac{8D^2}{v_e^4}
\eeq
The first term is the one from Eq.~\ref{drift_layer_position} due to the varying position of the photon interaction together with the average drift time, the second term is due to diffusion. The third term is an artefact of the assumption that a charge placed at $x=0$ can diffuse into a region of $x<0$, so the term does not vanish even for $w=0$. In a realistic implementation of a conversion layer, the region of $x<0$ in Fig.~\ref{drift_layer1} will represent a region where the electric field drops sharply to low values. The electron might spend a rather long time in this region before moving back into the high field region or it might even get lost. We therefore count the cases where the electron moves to this area as additional small inefficiency and neglect the term. In practice, photons interacting at the boundary of the conversion layer will cause a long tail in the time distribution and it becomes a practical question whether to include the tails in the calculation of the time resolution or count them as inefficiency.  In the limit of large and small values of $l_a/w $ the time resolution approximates to
\beq
     \sigma_t ^2 =  \frac{T^2}{12} + \frac{DT}{v_e^2}   \quad l_a  \gg w 
      \qquad     
      \sigma_t ^2 =  T^2 \frac{l_a ^2}{w^2}  +  \frac{2DT}{v_e^2}  \quad  l_a  \ll w 
\eeq
We assume $D=35$\,cm$^2/$s for electrons in silicon. For electric fields in excess of $5\times10^4$\,V/cm the electron velocity is close to  saturation and we have $D/(v_{sat}^e)^2=0.35$\,ps. For a conversion layer of $w=1/10/100\,\mu$m the saturated drift time is $T \approx 10/100/1000$\,ps.  \\
For $\l_a \gg w$ the probability for the photon conversion position becomes uniform across $w$, diffusion is negligible and the time resolution is $\sigma_t = T/\sqrt{12} \approx 2.89/28.9/289$\,ps for $w=1/10/100\,\mu$m. For $\l_a \ll w$ the photon conversion point is always close to $x=0$, diffusion will dominate and the time resolution is equal to $\sigma_t = \sqrt{2DT}/v_e \approx 1.6/8.37/26.46$\,ps for $w=1/10/100\,\mu$m.
\\ \\
If the illumination takes place from the 'right' side as indicated in Fig.~\ref{drift_layer1}b, the time resolution becomes
\beq
       \sigma_t^2 =  T^2 \left(\frac{l_a ^2}{w^2} - \frac{1}{4  \sinh[w/2l_a ]^2}\right) +T\frac{2D}{v_e^2}\left( \frac{l_a }{w}+\frac{1}{1-e^{w/l_a }} \right) + \frac{8D^2}{v_e^4}
\eeq
This expression differs from the previous one only by the diffusion term, because the distance between the conversion point and $x=w$ is now different. In the limit of large and small values of $l_a$ we have 
\beq
     \sigma_t ^2 =  \frac{T^2}{12} + \frac{DT}{v_e^2}  \quad l_a  \gg w 
      \qquad     
      \sigma_t ^2 =  T^2 \frac{l_a ^2}{w^2}  +  \frac{2DT}{v_e^2} \frac{l_a}{w} \quad  l_a  \ll w 
\eeq
For $l_a \gg w$ the expression is equal to the one from above, while for $l_a \ll w$ the variance goes to zero because the conversion point is close to $x=w$. The expression does of course not apply if the absorption length $l_a$ is of the same order or smaller than the gain layer thickness $d$, because the photons will interact directly in the gain layer.

\newpage


\section{Electron-hole avalanches and breakdown, average signal \label{avalanches}}

\begin{figure}[h]
\begin{center}
     \epsfig{file=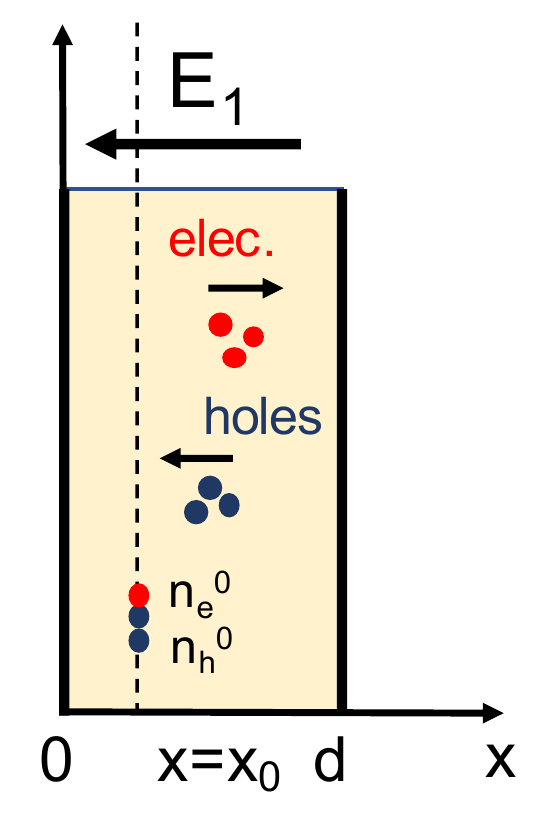, height=5cm}
\caption{The primary electrons and holes deposited at $x=x_0$ are multiplying, which results in a diverging avalanche in case the electric field $E_1$ is above the breakdown limit. }
\label{gain_layer}
\end{center}
\end{figure}
An electron drifting inside the conversion layer will move to the gain layer and trigger an avalanche starting from $x=0$. Alternatively a photon can convert inside the gain layer and the e-h pair at position $x=x_0$ will trigger the avalanche.  To cover both situations, we treat the general case where $n^0_e$ electrons and $n^0_h$ holes are deposited at $x=x_0$ at time $t=0$, as shown in Fig.~\ref{gain_layer}. \\
To derive equations describing the avalanche, we allow for general position-dependent electric fields $E_1(x)$. With the field orientated as shown in Fig.~\ref{gain_layer}, electrons move to the right and holes move to the left with velocities $v_e(x), v_h(x)$. The probability for an electron to create an e-h pair when travelling a distance $dx$ is $\alpha(x)dx$ while the probability for a hole to produce an e-h pair over distance $dx$ is $\beta(x) dx$, where $\alpha(x)$ and $\beta(x)$ are called the impact ionization coefficients.  The values for silicon (Fig.~\ref{silicon_parameters}) are reported in \cite{overstraeten_man} with the parameters listed in Appendix A. Since $1/\alpha$ and $1/\beta$ refer to the average distance that an electron or a hole has to travel in order to produce one additional e-h pair, we see that only for fields in excess of $2{-}3\times 10^5$\,V/cm there is an appreciable probability to provoke an avalanche in a few $\mu$m of silicon.
\begin{figure}[h]
\begin{center}
a)
     \epsfig{file=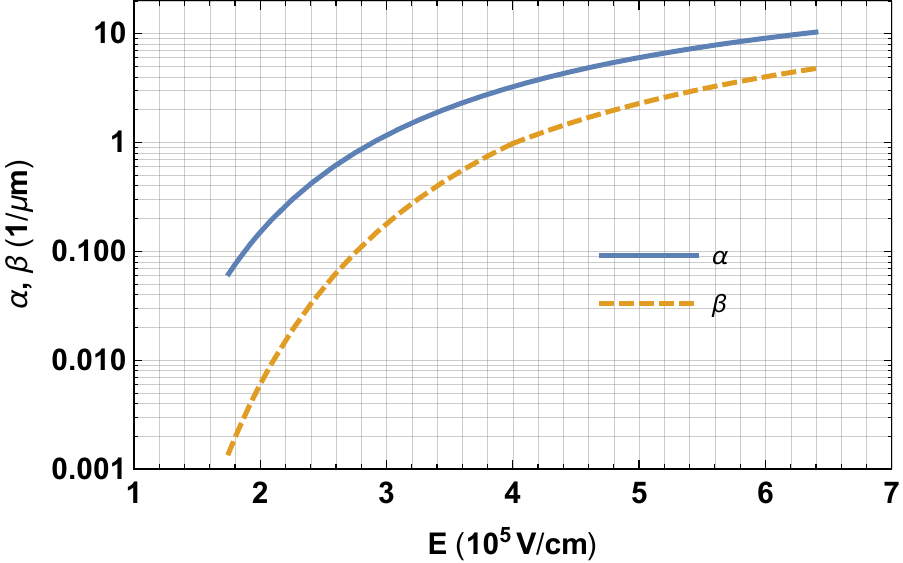, width=6.4cm}
b) 
      \epsfig{file=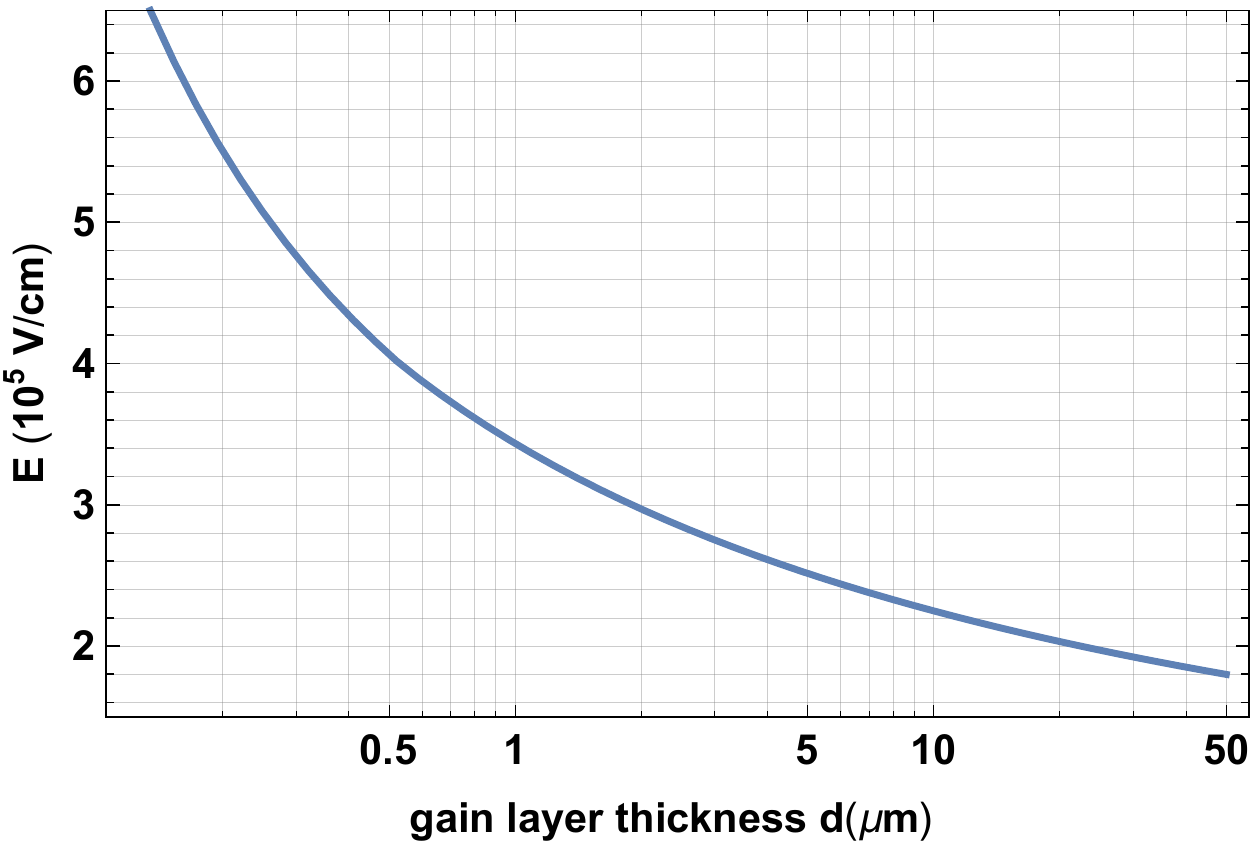, width=6cm}
\caption{ a) Impact ionization  coefficient $\alpha$ for electrons and $\beta$ for holes as a function of the electric field. b) Minimum electric field value provoking breakdown for a gain layer with constant electric field across a given thickness $d$. } 
\label{silicon_parameters}
\end{center}
\end{figure}
Fig.~\ref{MC} shows a Monte Carlo (MC) simulation of a few avalanches starting with a single electron. After some initial fluctuations the  avalanche just grows exponentially. There is also a finite probability that no diverging avalanche develops.
\begin{figure}[h]
\begin{center}
\epsfig{file=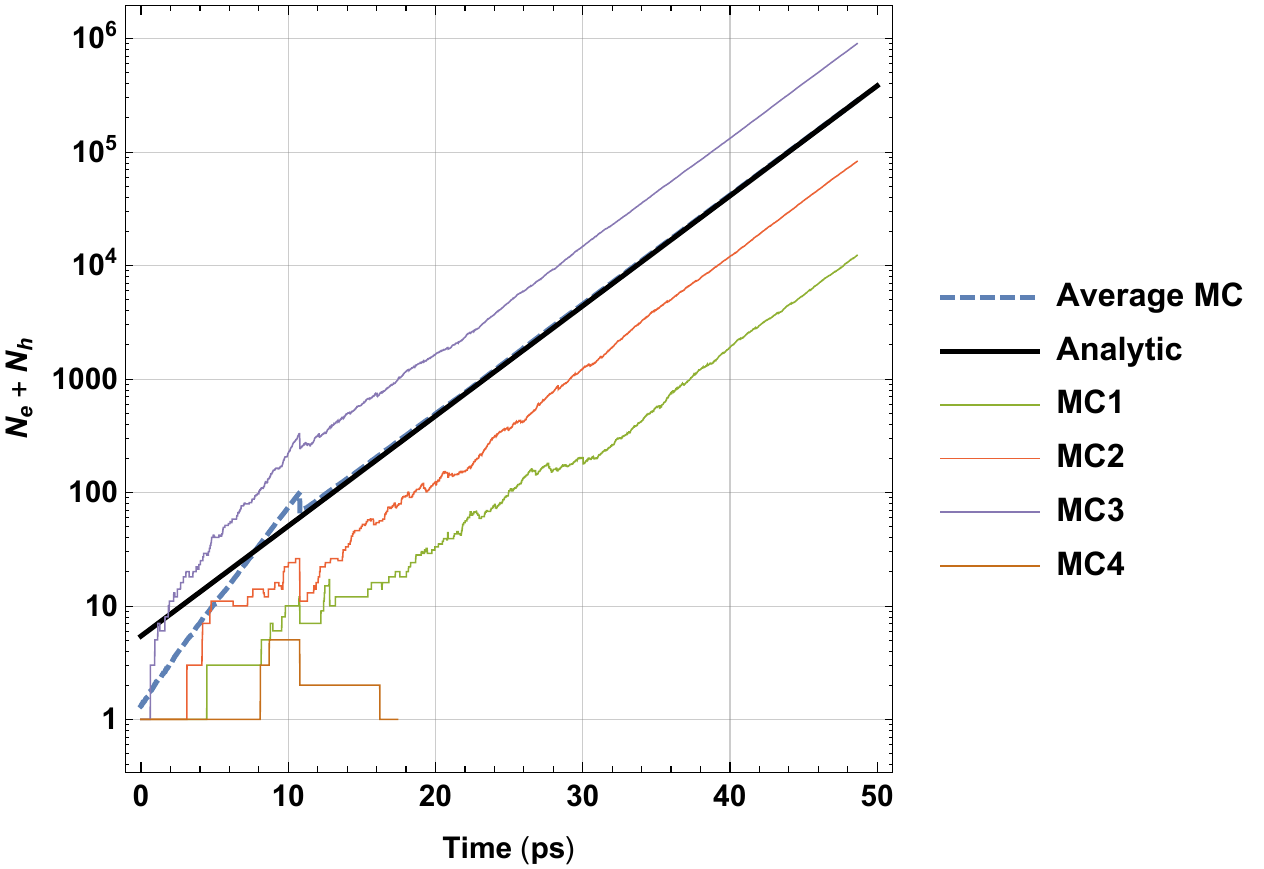, width=9cm}
\caption{Monte Carlo simulation for an electron-hole avalanche starting with a single electron at $x_0=0$ for a $1\,\mu$m diode at a field of 4\,V/$\mu$m. After some initial fluctuations the number of charge carriers increases exponentially. The dominant term from Eq.~\ref{tot_number_dominant_term} approximates the average signal extremely well for times $t>d/v^* \approx 11$\,ps. When the avalanche is still small there is a finite probability that no breakdown occurs, as is the case for the avalanche MC4.}
\label{MC}
\end{center}
\end{figure}
\\
We denote as $n_e(x,t)dx$  and $n_h(x, t)dx$ the average number of electrons and holes between position $x$ and $x+dx$ at time $t$ (note that this is different from the notation used in \cite{philipp_paper}). These charge densities result in local average current densities of $j_e(x,t) = v_e(x) n_e(x,t)$ and $j_h(x,t) = -v_h(x) n_h(x,t)$. By the continuity equation $\de j / \de x + \de n/\de t = \sigma$, with $\sigma$ the generation rate,  we therefore have 
\bea \label{average_equations} 
       \frac{\partial  n_e(x, t)}{\partial t}  + \frac{\partial  v_e(x) \,  n_e(x, t)}{\partial x} &=&
         \alpha(x) v_e(x) n_e(x, t) + \beta(x) v_h(x) n_h(x, t)   \\
       \frac{\partial  n_h(x, t)}{\partial t}  -\frac{\partial  v_h(x) \, n_h(x, t)}{\partial x} & = &
        \alpha(x) v_e(x)  n_e(x, t) + \beta(x)  v_h(x) n_h(x, t)  \no
\eea
The fact that electrons move to the left and holes move to the right gives the boundary conditions
\beq 
     n_e(0, t) = 0 \qquad  n_h(d, t)=0
\eeq
Since Eq.~\ref{average_equations}  represents a set of linear equations we can use the Ansatz $n_e(x, t)= f(x)e^{S t}$ and $n_h(x, t) = g(x)e^{S t}$ and we find
\bea \label{gamma_value_equation}
    S f(x)+ [v_e(x) f(x)]'  & = & \alpha (x) v_e(x) f(x) + \beta (x) v_h(x)  g(x) \\
     S g(x)  - [v_h(x) g(x)]'  & = &   \alpha  (x) v_e(x) f(x) + \beta (x) v_h(x)  g(x)  \nonumber
\eea
with $f(0) = 0$ and $g(d) =0$. The multiplication of electrons and holes can lead to a finite amount of total charge in the avalanche ($S<0$) or it can diverge and cause breakdown  ($S >0$). The boundary between the two regimes is at $S =0$, so by setting $S=0$ in the above equations and solving them with the given boundary conditions we find the breakdown condition (Appendix  B)
\beq \label{breakdown_condition} 
      \int_0^d \alpha(x) \, \exp \left[ -\int_0^x(\alpha(x')-\beta(x')) dx' \right] dx =1   
\eeq
The breakdown condition is independent of $v_e(x)$ and $v_h(x)$. This relation is usually derived by evaluating the point at which the gain for a constant current injected into the gain layer diverges \cite{stillman_wolfe}. Evaluating the breakdown equation for constant $\alpha$ and $\beta$  implies that breakdown occurs if 
\beq
    d>\frac{1}{\alpha-\beta} \ln \frac{\alpha}{\beta} 
\eeq
The electric field at which breakdown takes place for a gain layer with a given thickness $d$ is shown in Fig.~\ref{silicon_parameters}b. For general electric field profiles, Eqs.~\ref{average_equations} can be efficiently solved with numerical methods. The solution can be given in analytical form for constant values of $\alpha, \beta, v_e, v_h$ and is derived in \cite{philipp_paper}. 
It is represented as an infinite sum of exponential terms with (generally complex valued) time constants. At least one time constant is guaranteed to be real-valued. The largest real-valued time constant defines the long-term behaviour of the avalanche. Above the breakdown limit, this term determines the rate of exponential growth of the avalanche. Starting with  
 $n_e^0$ electrons and $n_h^0$ holes at position $x_0$ at time $t=0$, it reads
\beq
     n_e(x, t)   =   \frac{1}{d}\,a(x)  \left[ n_e^0u_e(x_0) + n_h^0u_h(x_0)  \right] \,e^{\gamma v^* t} \qquad 
   n_h(x, t)   =     \frac{1}{d} \, b(x)    \left[ n_e^0u_e(x_0) + n_h^0u_h(x_0)  \right]\,e^{\gamma v^* t}
\eeq
with
\bea \label{u_functions}
      u_e(x_0) & = & 
      e^{-a_1 x_0} \sin (k-k x_0/d ) \label{ue_function} \\
      u_h(x_0)  & = &      \, \frac{e^{-a_1 x_0} }{\alpha d}  \label{uh_function}
   \left[
      k \cos\left(k-k x_0/d \right) 
     +\lambda_1 \sin (k- k x_0/d )  
      \right] \\
    a(x) & = &  \frac{2 v_h k e^{a_1 x} \sin \left( k\frac{x}{d} \right)}{(v_e+v_h)(1+\lambda_1)\sin k}   \label{ax_function}\\
     b(x) & = &  \frac{2 k  v_e e^{a_1 x} \left[  k \cos \left( k\frac{x}{d} \right)+ \lambda_1\sin \left( k\frac{x}{d} \right) \right]  }{(v_e+v_h)\beta d (1+\lambda_1)\sin k}    \label{bx_function}           
\eea
and the constants $v^*, a_1, \gamma, k$ are defined by 
\bea
     v^* & = &  \frac{2 v_e v_h}{v_e+v_h}   \label{vstar} \\
     a_1 & = & \frac{\alpha v_e - \beta v_h}{v_e+v_h} + \frac{v_e-v_h}{v_e+v_h} \frac{1}{d}\ \, \lambda_1 \\
     \gamma & = & \frac{\alpha+\beta}{2} + \frac{\lambda_1}{d} \label{gamma_equation} \label{gamma} \\
    k & = & \sqrt{ \alpha\beta d^2 -\lambda_1^2 } 
\eea
The parameter $\lambda_1$ is the largest real solution of the equation 
\beq  \label{lambda_equation}
       \lambda_1+ \sqrt{ \alpha\beta d^2 - \lambda_1^2} \cot\sqrt{\alpha\beta d^2 - \lambda_1^2}= 0
\eeq
It holds that $-\infty < \lambda_1 < d \sqrt{\alpha \beta}$. For $\lambda_1 < -  d\sqrt{\alpha \beta}$ the constant $k$ will become imaginary, which will still lead to real valued expressions for $n_e(x, t)$ and $n_h(x, t)$ with $\sin, \cos$ becoming $\sinh, \cosh$. The functional form of $a(x)$ and $b(x)$ as well as the equation for $\gamma$ were already derived in \cite{holway}. Fig.~\ref{ueuh}a shows the functions $u_e(x_0), u_h(x_0), u_e(x_0)+u_h(x_0)$ that determine how the average growth of the avalanche depends on the position of a primary electron, hole or e-h pair. They are the mirror images of the functions $a(x), b(x)$ from Eqs.~\ref{ax_function}, \ref{bx_function} that determine the distribution of the electrons and holes inside the gain layer  \cite{philipp_paper}. 
\\ \\
The parameter $\gamma$ defines the exponential growth of the avalanche and is shown in Fig.~\ref{ueuh}b. It has the following properties:
\bea
\gamma & = & 0 \qquad \qquad  \qquad  \quad  \, \, \alpha\beta d^2 = \frac{\alpha \beta}{(\alpha-\beta)^2} \ln^2\frac{\beta}{\alpha} \le 1 \\
     \gamma &  =  &  \frac{\alpha+\beta}{2} - \sqrt{\alpha \beta} \qquad \alpha\beta d^2=1 \\
     \gamma &  =  &  \frac{\alpha+\beta}{2} \phantom{ + \sqrt{\alpha \beta} } \qquad \alpha\beta d^2= \frac{\pi^2}{4}  \approx 2.47 \\
      \gamma_{max} &  =  &  \frac{\alpha+\beta}{2}  + \sqrt{\alpha \beta} \qquad \alpha \beta d^2\rightarrow \infty
\eea
\begin{figure}[h]
\begin{center}
a)
\epsfig{file=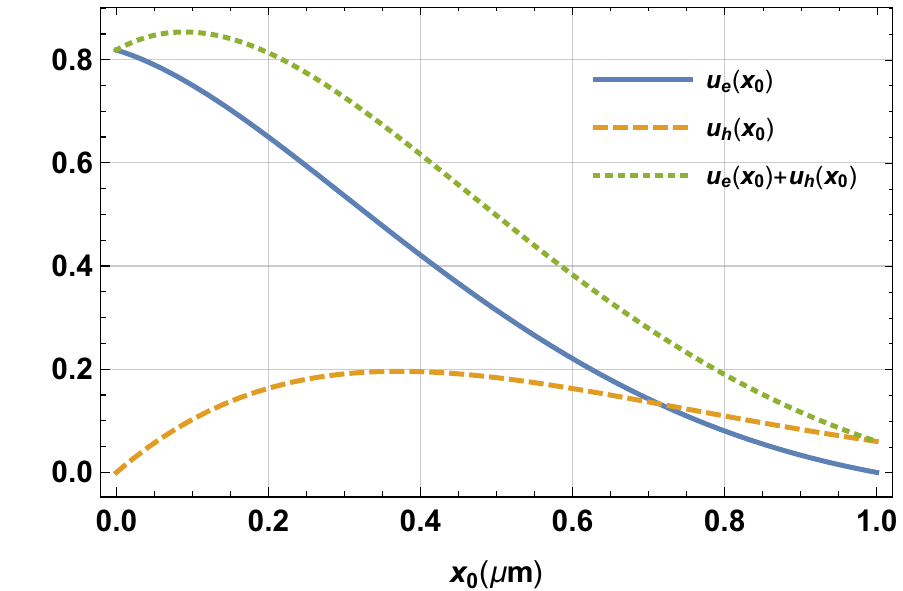, width=6cm}
b)
\epsfig{file=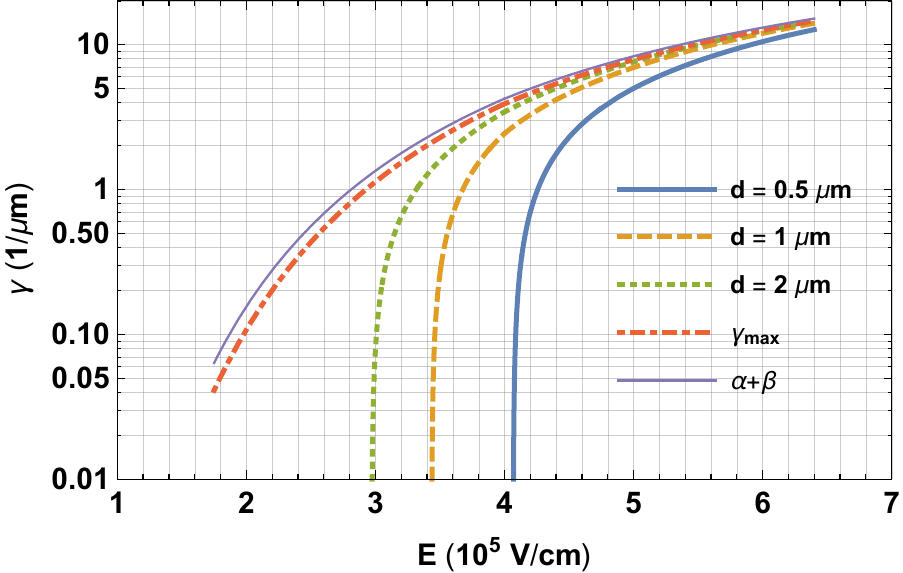, width=6cm}
\caption{a) The functions $u_e(x_0), u_h(x_0), u_e(x_0)+u_h(x_0)$ from Eqs.~\ref{ue_function}, \ref{uh_function} that determine how the average growth of the avalanche depends on the position of a primary electron, hole or e-h pair. The values are for a $1\,\mu$m gain layer at $E_1=4.5\times10^5$\,V/cm.
 b) $\gamma$ as a function of electric field in silicon for different values of the gain layer thickness $d$. At the breakdown limit we have $\gamma=0$. For higher electric fields $\gamma$ quickly approach $\gamma_{\max}$. }
\label{ueuh}
\label{inverse}
\end{center}
\end{figure}
\\
The total number of electrons and holes is then given by $N=\int_0^d n(x,t) dx$ and evaluates to
\beq \label{tot_number_dominant_term}
     N_e(t)  =   B_e   \,\left[ n_e^0u_e(x_0) + n_h^0u_h(x_0)  \right] \,e^{\gamma v^* t}  \qquad 
      N_h(t)  =   B_h \,  \left[ n_e^0u_e(x_0) + n_h^0u_h(x_0)  \right] \,e^{\gamma v^* t} \qquad 
\eeq
where we have
\bea 
    B_e& = &   
    \frac{2kv_h 
     \left[
    e^{a_1 d} (a_1 d-k\cot k)+k\csc k
    \right]
    }{(v_e+v_h)(a_1^2d^2+k^2)(1+\lambda_1)} \label{Be_coefficients} \\
     B_h & = & 
       \frac{2kv_e 
    \left[
    e^{a_1 d} (k^2+a_1d\lambda_1+k(a_1 d-\lambda_1)\cot k)
    +k (\lambda_1-a_1 d) \csc k
    \right]
    }
    {(v_e+v_h)\beta d (a_1^2d^2+k^2)(1+\lambda_1)} \label{Bh_coefficients}
\eea
The total induced current becomes 
\bea \label{average_current} 
      I(t)  &   = &    e_0 \frac{E_w}{V_w}  \left[v_eN_e(t)+v_h N_h(t) \right]  \\
      & = &  e_0\frac{E_w}{V_w}  \left(v_e B_e +v_h B_h \right)\left[ n_e^0u_e(x_0) + n_h^0u_h(x_0)  \right]  e^{\gamma v^* t}
\eea
where $e_0$ is the electron charge. Here we have assumed a constant weighting field $E_w/V_w$ in the region in which the charges are moving. For the single photon detection using the conversion layer we have only a single electron at $x_0=0$ and therefore $n_e^0=1, n_h^0=0$ and the expression is
\beq
    I(t)  = e_0\frac{E_w}{V_w}  \left[ v_e B_e +v_h B_h \right] \sin k  \, e^{\gamma v^* t}
\eeq 
Assuming a velocity $v_e \approx v_h \approx v_{sat} \approx 0.1\,\mu$m/ps and a weighting field of $E_w/V_w = 1/d=1/(1\,\mu$m), the current corresponding to $10^5$ charges at $t=44$\,ps in Fig.~\ref{MC} is 1.6\,mA.


\newpage

\section{Efficiency \label{efficiency} }

\begin{figure}[h]
\begin{center}
a)
     \epsfig{file=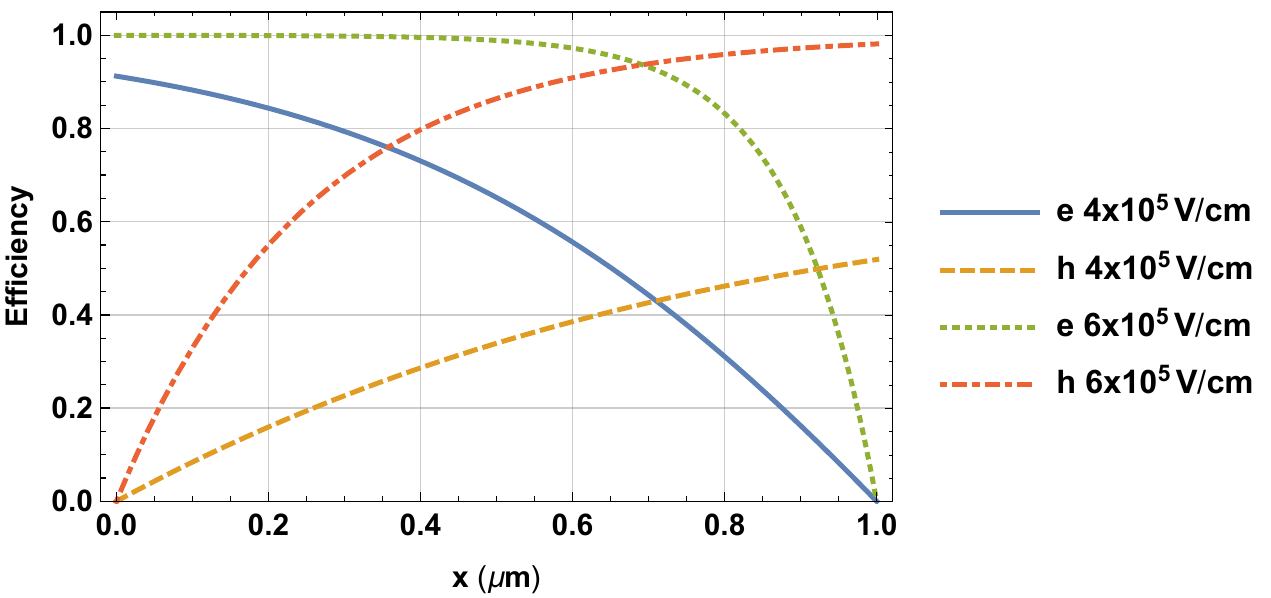, width=7cm}
     b)
          \epsfig{file=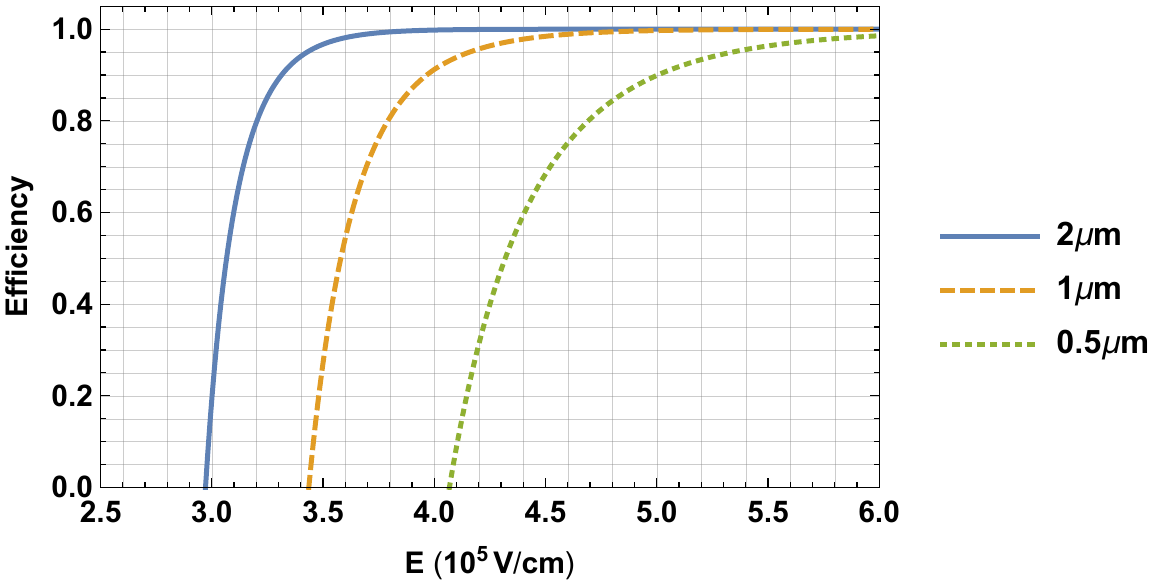, width=7cm}
\caption{a) Breakdown probability (efficiency) for a single electron and a single hole deposited at position $x$ inside a gain layer of $d=1\,\mu$m for two values of the electric field. b) Breakdown probability $p_0$ (efficiency) for a single electron placed at $x=0$ for different values of the gain layer thickness $d$.}
\label{eff1}
\end{center}
\end{figure}

In this section we calculate the probabilities $P_e(x)$ and $P_h(x)$ for a single electron or a single hole placed at position $x$ in the gain layer to cause breakdown. We follow  \cite{triggering_phenomena}  to establish the equations for these quantities. We start by considering a single electron at position $x-dx$ which moves in positive $x$-direction in the applied electric field. The probability for it to create a diverging avalanche is $P_e(x-dx)$. Between $x-dx$ and $x$ two things can happen. 1) With a probability of $(1-\alpha dx)$ there is no multiplication of the electron and then the electron at position $x$ creates a diverging avalanche or 2) the electron is multiplying over the distance $dx$ and at least one of the two electrons and the hole create breakdown. This can be written as
\beq
     P_e(x-dx) = (1-\alpha dx)P_e(x) + \alpha dx \left[  1-(1-P_e(x))^2(1-P_h(x) \right]
\eeq 
Writing the corresponding equation for $P_h(x)$ and expanding for small $dx$ gives 
\bea \label{turnon_equations}
   \frac{d \, P_e(x)}{d\,x} & = &  -\alpha(x) [1-P_e(x)] \left[ P_e(x) + P_h(x) - P_e(x) P_h(x) \right] \\
      \frac{d \, P_h(x)}{d\,x} & = &  \phantom{+}\beta(x) [1-P_h(x)] \left[ P_e(x) + P_h(x) - P_e(x) P_h(x) \right] \nonumber
\eea
Provided $\alpha(x)$ and $\beta(x)$ are known, these equations can be integrated with the boundary conditions $P_e(d) = 0$ and $P_h(0) = 0$. Following  \cite{mcintyre_breakdown} we define $P(x) = P_e(x) + P_h(x) - P_e(x) P_h(x)$ and by differentiating this expression and using Eqs.~\ref{turnon_equations} we have
\beq
    \frac{d \, P(x)}{ d\, x} = -(\alpha-\beta) P(x)[1-P(x)]
\eeq
We use the boundary condition $P(0)=P_e(0)=p_0$, with $p_0$ still to be determined, giving the solution
\beq \label{breakdown_ehpair} 
      P(x) = \frac{p_0}{p_0+(1-p_0) \exp \left[ \int_0^x (\alpha(x')-\beta(x'))dx'\right]}
\eeq
Since $P(x)$ can be written as $P(x) = 1- [(1-P_e(x))(1-P_h(x))]$ we see that $P(x)$ refers to the breakdown efficiency of a single e-h pair, which we will use later for the efficiency calculation for MIPs. Knowing $P(x)$ and using $P_h(d)=0$ we can integrate Eqs.~\ref{turnon_equations} and have 
\beq
     P_e(x) = 1-\exp \left( - \int_x^d \alpha(x') P(x') dx'  \right)  \qquad  P_h(x) = 1-\exp \left( - \int_0^x \beta (x') P(x') dx'  \right)
\eeq
And finally $P_e(0)=p_0$ gives the equation that allows us to determine $p_0$ 
\bea \label{p0_equation}
          p_0 = 1 - \exp \left( - \int_0^d\frac{p_0\,  \alpha(x')}{p_0+(1-p_0) \exp \left[ \int_0^{x'} (\alpha(x'')-\beta(x''))dx''\right]} dx'  \right) 
\eea
In general this equation can only be evaluated numerically. This equation also reveals again the breakdown condition. Close to the threshold of breakdown the value of $p_0$ will be small. 
For small values of $p_0$ the expression $1-\exp \left[ -p_0/(p_0+(1-p_0)e^v)) \right]$ is approximated by  $p_0 e^{-v}+O(p_0^2)$, so the above relation turns into the breakdown condition of Eq.~\ref{breakdown_condition}.
\\ 
For constant $\alpha$ and $\beta$ the above expressions evaluate to 
\bea
      P(x) & = & \frac{p_0}{p_0+[1-p_0]e^{(\alpha-\beta)x} }  \label{breakdown_eh_constant}   \\
         P_e(x) & = & 1 - e^{-\alpha (d-x)}\left[ \frac{(1-p_0)e^{(\alpha-\beta)d}+p_0 }{(1-p_0)e^{(\alpha-\beta)x}+p_0} \right]^\frac{\alpha}{\alpha-\beta}   \label{breakdown_e_constant} \\
         P_h(x) & = & 1-e^{-\beta x}\left[ (1-p_0)e^{(\alpha-\beta)x}+p_0 \right]^\frac{\beta}{\alpha-\beta}  \label{breakdown_h_constant} 
\eea
Eq.~\ref{p0_equation}  that determines $p_0$ reads as 
\beq
            e^{-(\alpha-\beta)d} =  \frac{1}{p_0}\left[ (1-p_0)^{1-\frac{\beta}{\alpha}} -(1-p_0) \right]
\eeq
Fig.~\ref{eff1} shows a few examples.


\section{Time resolution \label{time_resolution}}

The statistics of electron-hole avalanches and the resulting contribution to the time resolution are discussed in detail in \cite{philipp_paper}, where the problem is treated by using the theory of continuous-time Markov processes.  

\subsection{Avalanches in absence of boundaries}

For the case of an e-h avalanche in absence of any boundaries and for a constant electric field, an alternative approach can be used to derive the avalanche fluctuations \cite{legler} that does not require the formalism developed in \cite{philipp_paper} and which is closely related to the arguments resulting in Eq.~\ref{turnon_equations}. We define $p_e(n, m, \Delta)$ to be the probability to find $n$ electrons and $m$ holes at time $t_0+\Delta$ for an avalanche starting with a single electron at $t_0$. For an avalanche starting at $t=0$, there are two ways to reach this state at the later time $t+dt$. First, the initial electron does not multiply in the first small time interval $[0, dt]$ (with probability $1-\alpha v_e dt$), but then produces $n$ electrons and $m$ holes during the subsequent time interval $[dt, t+dt]$. This happens with a probability $p_e(n, m, t)$. Second, the electron already multiplies in the interval $[0, dt]$ (with probability $\alpha v_e dt$) and the resulting two electrons and one hole multiply into $n$ electrons and $m$ holes during $[dt, t+dt]$. This is written as
\bea \label{penmt}
        p_e(n, m, t+dt) & = &  (1-\alpha v_e dt)  p_e(n, m, t) \\
        & + & \alpha v_e dt \sum_{i=1}^n \sum_{j=1}^i \sum_{r=1}^m \sum_{s=1}^r p_e(n-i-j,m-r-s,  t)p_e(i,r,t)p_h(j,s,t)  
\eea
where $p_h(n, m, \Delta)$ is the probability that an avalanche starting with a single hole at time $t_0$ produces $n$ electrons and $m$ holes at $t_0 + \Delta$. Writing the corresponding equation for $p_h(n, m, t)$ and expanding for small $dt$ results in the equations defining $p_e(n, m, t)$ and $p_h(n, m, t)$. The equations have a structure similar to Eq.~\ref{turnon_equations}. Their solution is given in Appendix C and equal to the one derived in \cite{philipp_paper}. Having $n_e^0$ electrons and $n_h^0$ holes at $t=0$, the probability to have $n$ additionally created e-h pairs at time $t$ is 
\beq \label{p_additional_eh}
     p(n, t) = \frac{\Gamma(A+n)}{\Gamma (A) \Gamma (1+n) } \left(  \frac{1}{\nu (t)} \right)^A \left( 1 - \frac{1}{\nu (t)}  \right)^{n} \qquad
      \sum_{n=0}^\infty p(n, t)=1
\eeq
\beq
    A=\frac{n_e^0\alpha v_e+ n_h^0\beta v_h}{\alpha v_e+\beta v_h}  \qquad 
    \nu (t) = e^{\lambda_t t}   \qquad
    \lambda_t  = \alpha v_e+\beta v_h
\eeq
The average number of e-h pairs and the variance are
\beq
     \ov n(t) = [\nu(t)-1]A \qquad  \sigma_n^2(t) =  \nu(t) [ \nu (t)-1] A
\eeq
In the continuous approximation for $n$, which is valid for large $n$ (and therefore for large avalanches i.e.~late times)  we have
\beq
      p(n, t) =  \frac{n^{A-1}}{\Gamma (A) }\,\left(  \frac{A}{\ov n (t)} \right)^A e^{-n A/\ov n(t)} \qquad \int_0^\infty p(n,t) dn = 1
\eeq
with
\beq
       \ov n(t) = A e^{\lambda_t t} \qquad \sigma_n(t) =   \frac{1}{\sqrt{A}} \ov n(t) 
\eeq
\begin{figure}[h]
\begin{center}
a)
\epsfig{file=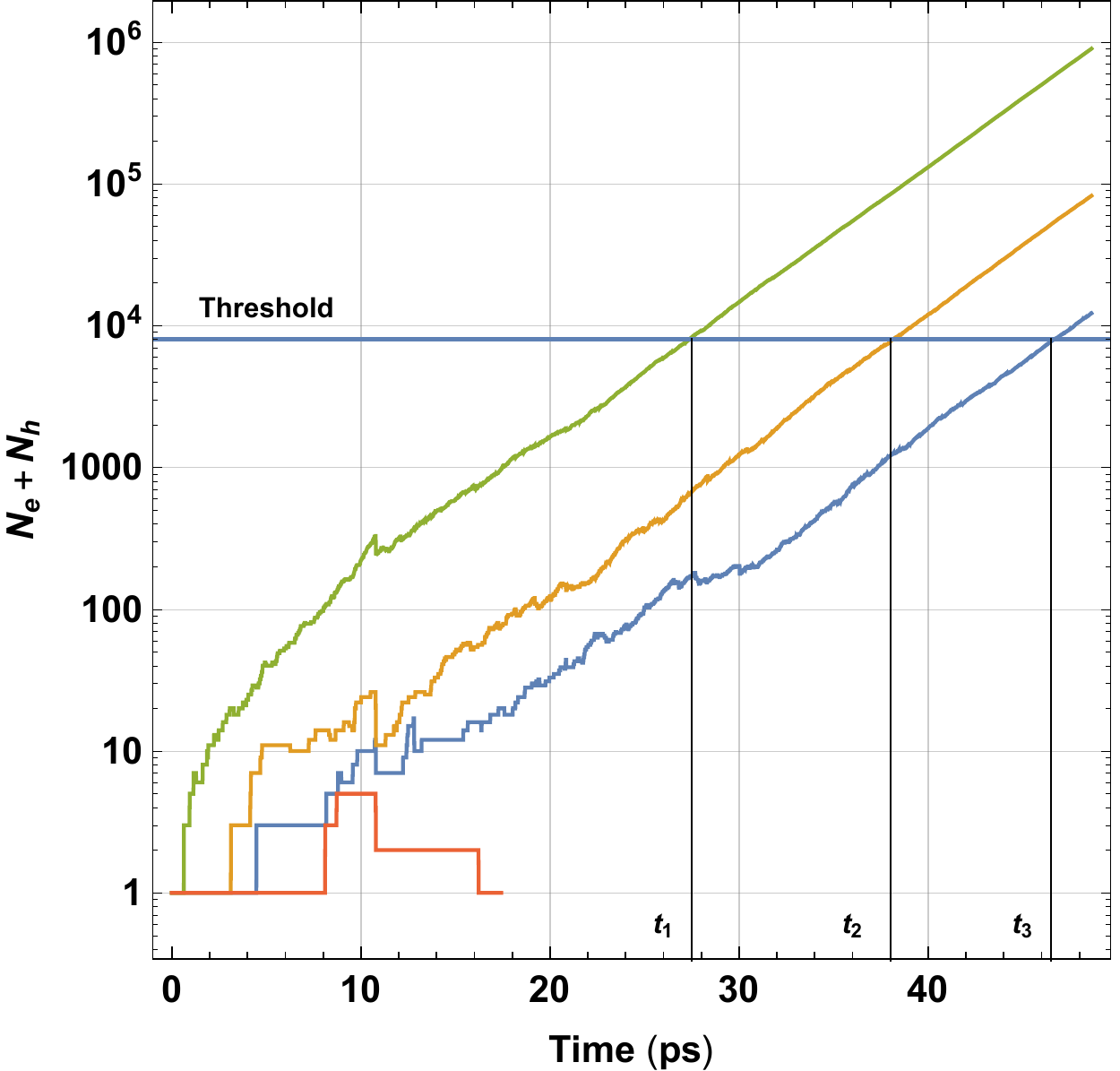, width=6cm}
b)
\epsfig{file=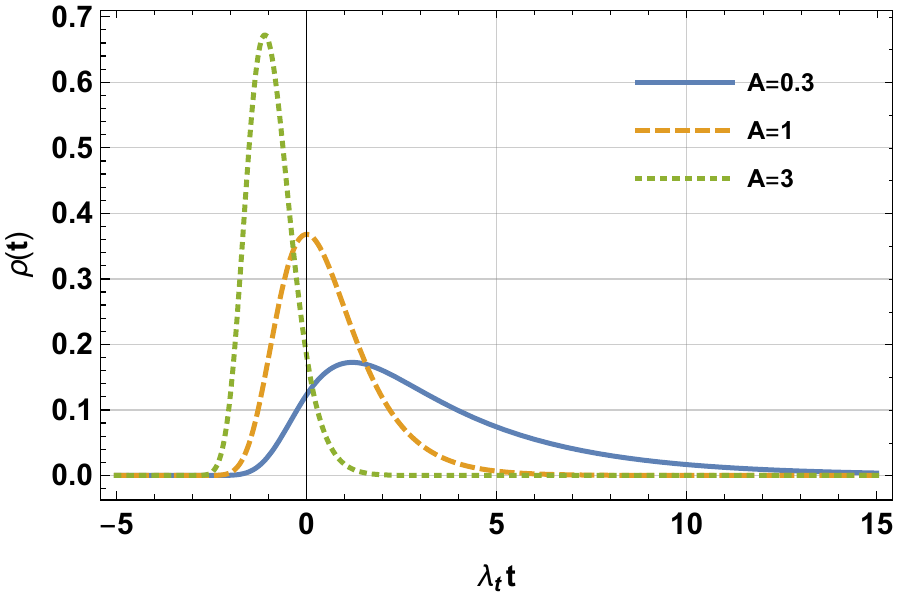, width=7cm}
\caption{a) A threshold is applied to the signal. The fluctuations of the avalanche size result in a fluctuation of the threshold crossing time, which defines the time resolution. b) Time response function $\rho(t)$ for different values of the parameter $A$. }
\label{threshold}
\end{center}
\end{figure}
In order to measure the 'signal time' we apply a threshold to the signal, which is proportional to the total number of charge carriers. Fig.~\ref{threshold}a shows how the avalanche fluctuations lead to fluctuations of the threshold crossing time, which determine the time resolution. The probability that the signal crosses the threshold of $n$ e-h pairs between time $t$ and $t+dt$, the so-called time response function, is given by
\beq
     \rho(n, t) dt =\lambda_t \,  \frac{\Gamma(1+A+n)}{\Gamma (A) \Gamma (1+n) } \left(  \frac{1}{\nu (t)} \right)^A \left( 1 - \frac{1}{\nu (t)}  \right)^{n} dt
     \qquad \int_0^\infty \rho(n, t)dt = 1
\eeq
The average threshold crossing time and its variance are given by
\beq
   \ov t =  \frac{1}{\lambda_t } \left[ \psi_0 (n+A+1) - \psi_0(A) \right]
   \qquad
    \sigma = \frac{1}{\lambda_t }  \sqrt{ \psi_1 ( A) - \psi_1(n+A+1) } 
\eeq
where $\psi_0(x)=d \ln \Gamma(z)/dz$ is the digamma function and $\psi_1(x)=d^2 \ln \Gamma(z)/dz^2$ is the trigamma function. For large numbers of $n$ the above time response function approximates to 
\beq
        \rho(n, t) =   \frac{\lambda_t}{\Gamma(A)}  \exp \left[ A \ln n -A \lambda_t  t - ne^{-\lambda_t  t} \right] 
\eeq 
and the average threshold crossing time and the time resolution approximate to
\beq \label{time_reso_approx}
   \ov t =  \frac{1}{\lambda_t } \left[ \log n- \psi_0(A) \right]
   \qquad
    \sigma = \frac{1}{\lambda_t }  \sqrt{ \psi_1 ( A)  } 
\eeq 
so the time resolution becomes independent of the threshold. This can be understood when looking at Fig.~\ref{threshold}a and it is a well-established fact for detectors like Resistive Plate Chambers \cite{RPC}, where avalanche fluctuations dominate the signal characteristics  \cite{RPC_mangiarotti, RPC_werner}: scaling the threshold by a constant $c_1$  will just shift the time response function by $\Delta t =( \ln c_1)/\lambda_t$ without altering its shape. If we are not interested in the absolute time of the threshold crossing but just the time variations we can arbitrarily set $n=1$ and have the time response function
\beq \label{time_response_continuous} 
     \rho(t) =   \frac{\lambda_t}{\Gamma(A)}  \exp \left[ -A \lambda_t  t - e^{-\lambda_t  t} \right] \qquad \int_{-\infty}^\infty \rho(t) dt = 1
\eeq
An example of the time response function for different values of the parameter $A$ is shown in Fig.~\ref{threshold}b. 
The function $\sqrt{\psi_1(x)}$ approaches $\sqrt{1/x+1/x^2}$ for small and large values $x$ and it is within 10\% in the entire range of $x>0$, as can be seen in Fig.~\ref{psi}a. The values of $\sqrt{\psi_1(A)}$ for a primary electron, a primary hole and a primary e-h pair are shown in Fig.~\ref{psi}b. \\
We see that when starting with a primary hole, the time resolution is significantly worse when compared to a primary electron or a primary e-h pair.  An electron travels on average a distance of $1/\alpha$ and a hole  travels on average a distance of $1/\beta$ before creating an additional e-h pair. Since $\alpha > \beta$ in silicon the holes will travel significantly longer than the electrons before triggering the breakdown and therefore the time fluctuations are larger. The geometries of SPADs and SiPMs produced from silicon are therefore built such that the electrons trigger the avalanche in the gain region. Charge carrier mobilities are quite different in other semiconductors and the arrangement of doping layers can therefore differ.
\begin{figure}[h]
\begin{center}
a)
\epsfig{file=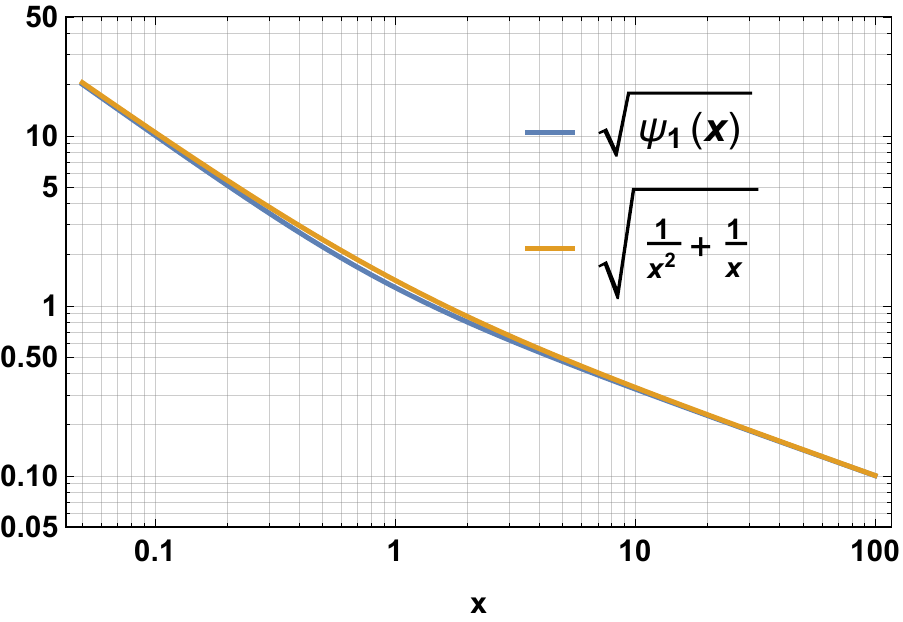, width=6cm}
b)
\epsfig{file=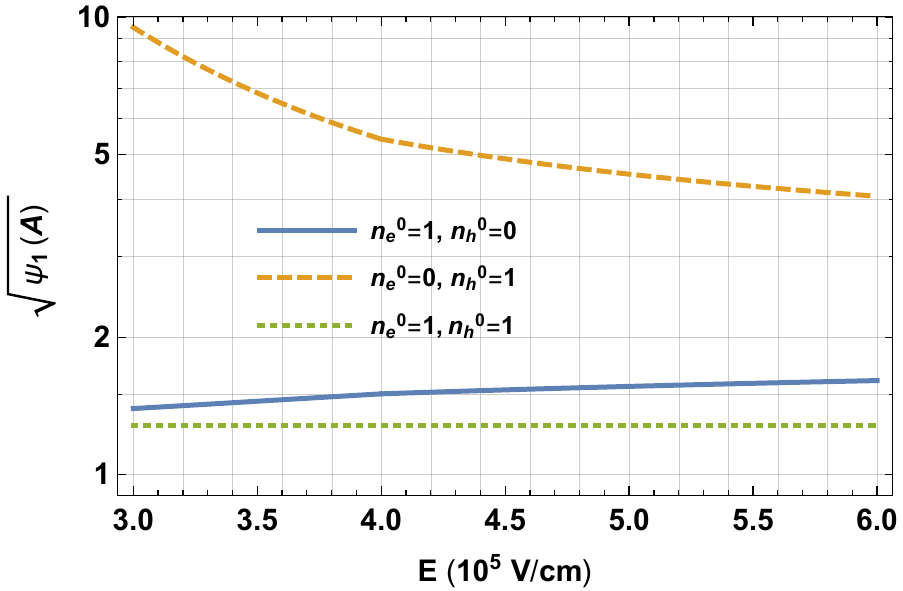, width=6.3cm}
\caption{a) The trigamma function $\psi_1(x)$. b) Values of $\sqrt{\psi_1(A)}$ for different initial conditions. Starting with a single e-h pair we have $A=1$ and  $\sqrt{\psi_1(1)}=\pi/\sqrt{6} \approx 1.28$. }
\label{psi}
\end{center}
\end{figure}

\clearpage

\subsection{Avalanches in a gain layer of finite thickness} 

Our central result in \cite{philipp_paper} is the conclusion that the finite thickness of the gain layer will to first order not affect the avalanche fluctuations but will only affect the average growth of the avalanche. This approximation works best if the primary charge carrier has the larger impact ionization coefficient, i.e.~if the primary charge in the gain layer is either an electron or an e-h pair for silicon. Then, the probability to have a total number of $n$ electrons and $m$ holes  at time $t$ in the gain layer, starting with $n_e^0$ electrons and $n_h^0$ holes at $x=x_0$ at time $t=0$, is 
\beq
      p(n, t) \approx  [1-\varepsilon(x_0)]\delta(n)  +\varepsilon(x_0)  \frac{n^{A-1}}{\Gamma (A) }\,\left(  \frac{A}{ N_e (t)} \right)^A e^{-n A/ N_e (t)} 
      \qquad m = \frac{B_h}{B_e} n 
      \qquad  A=\frac{n_e^0\alpha v_e+ n_h^0\beta v_h}{\alpha v_e+\beta v_h} 
      \qquad 
\eeq
with $N_e(t)$ from Eq.~\ref{tot_number_dominant_term} and $B_e, B_h$ are from Eqs.~\ref{Be_coefficients}, \ref{Bh_coefficients}. The number of electrons $n$ and the number of holes $m$ are considered continuous variables and taken to be fully correlated in this approximation (as shown in \cite{philipp_paper}, this is strictly true only at late times). The efficiency $\varepsilon(x_0)$ that the $n_e^0$ electrons and $n_h^0$ holes at $x{=}x_0$ trigger a diverging avalanche is
\beq
    \varepsilon(x_0) = 1-[1-P_e(x_0)]^{n_e^0}[1-P_h(x_0)]^{n_h^0}
\eeq
with $P_e(x_0)$ and $P_h(x_0)$ from Eqs.~\ref{breakdown_e_constant}, \ref{breakdown_h_constant}.
%
%
%
%
%
%
The corresponding time response function in reference to Eq.~\ref{time_response_continuous} is then  
\beq
        \rho(t) \approx   \frac{ \gamma v^*\,h^A}{\Gamma (A) }   \exp \left[ -A \gamma v^* t -h e^{-\gamma v^*  t}\right] 
\eeq
with
\beq
        h = \frac{A \, \varepsilon (x_0)} {B_e \left[ n_e^0u_e(x_0) + n_h^0u_h(x_0)  \right] } 
\eeq
Here we have divided $N_e(t)$ by the efficiency in order to account for the fact that only diverging avalanches are crossing the threshold. The corresponding time resolution due to the avalanche fluctuations, when keeping $n_e^0, n_h^0, x_0$ constant, is
\beq \label{timeres_bounded_eqn_approx}
        \sigma = \frac{\sqrt{\psi_1 ( A)}}{\gamma v^*}
\eeq
where $\gamma$ and  $v^*$ are from Eq.~\ref{vstar} and Eq.~\ref{gamma_equation}. The value of $1/\gamma v^*$ for different values of the thickness $d$ of the gain layer is shown in Fig.~\ref{inverse}a. 
\begin{figure}[h]
\begin{center}
a)
\epsfig{file=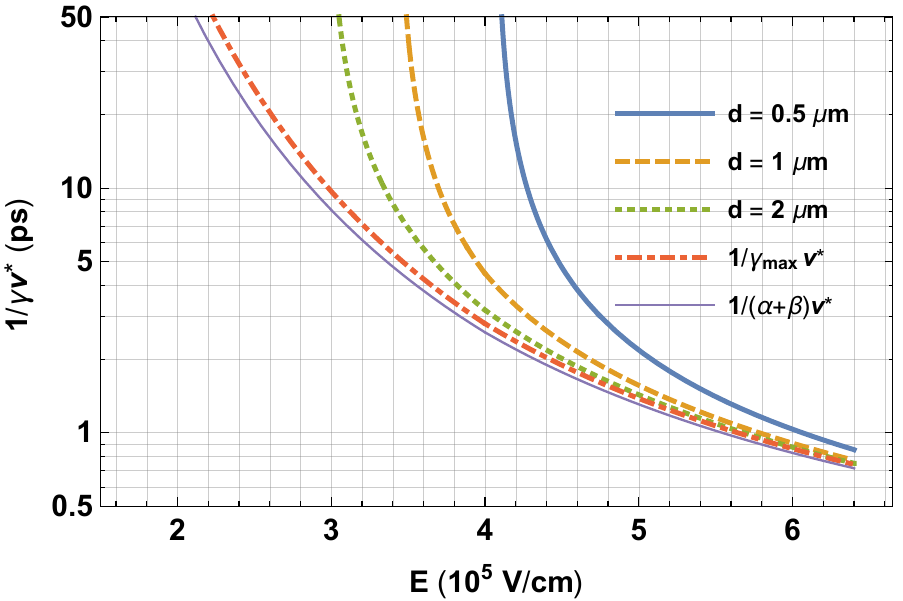, width=5.5cm}
b)
 \epsfig{file=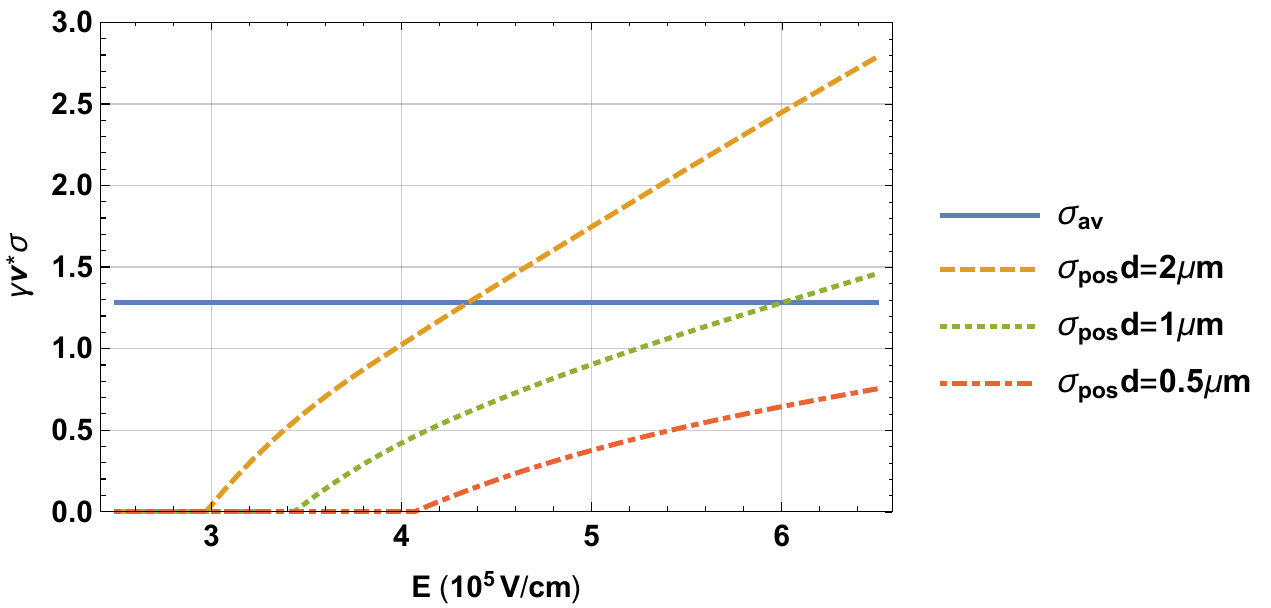, width=7.8cm}
\caption{a) The factor $1/\gamma v^*$ for different values of the gain layer thickness. b) Contributions to the time resolution for a photon interacting in the gain layer. The value of $\pi/\sqrt{6}$ indicated by the horizontal line is due to avalanche fluctuations and the other lines are due to fluctuations of the position of the photon conversion assuming a uniform distribution. The two components have to be added in squares, so we see that for thick gain layers and high fields the position dependence starts to dominate over the avalanche fluctuations. }
\label{inverse}
\end{center}
\end{figure}
\\
For a photon interacting in the conversion layer (Fig.~\ref{spad_scheme}b), the electron arriving at the gain layer will therefore start an avalanche from $x_0=0$ and the time resolution contribution of the gain layer will be 
\beq 
              \sigma \approx \frac{\sqrt{\psi_1 ( A)}}{\gamma v^*} \qquad A= \frac{\alpha v_e}{\alpha v_e + \beta v_h} 
\eeq                 
As seen in Fig.~\ref{psi} we have $\sqrt{\psi_1 ( A)} \approx 1.5$ in the entire electric field range, so e.g.~for a $1\,\mu$m SPAD at $4\times 10^5\,$V/cm the contribution to the time resolution will be around 6\,ps. For higher fields a time resolution of 1\,ps should theoretically be achievable. \\ 
In case the photon interacts inside the gain layer and produces an e-h pair, the varying position of the photon interaction will also contribute to the time resolution. Having one e-h pair at position $x_0$ as shown in Fig.~\ref{spad_scheme}a we have $n_e^0=n_h^0=1$ and therefore $A=1$, $\varepsilon(x_0) = P(x_0)$ from Eq.~\ref{breakdown_eh_constant} and the time response function is
\beq \label{hx0}
         \rho(t, x_0) =  \gamma v^*\,h(x_0)  \exp \left[ - \gamma v^* t -h(x_0) e^{-\gamma v^*  t}\right] \qquad h(x_0) = \frac{P(x_0)} {  B_e \left[ u_e(x_0) + u_h(x_0)  \right] } \qquad 
\eeq
The probability of conversion at position $x_0$ is given by 
\beq \label{conversion_probability2} 
     p_1(x_0)dx_0 = \frac{1}{1-e^{-d/l_a}}\frac{1}{l_a} e^{-x_0/l_a} dx_0
\eeq
and the time response function including the fluctuation of the conversion point is therefore defined by 
\beq
    \ov \rho(t) =  \int_0^d  p_1(x_0) \rho(t, x_0) dx_0 
\eeq
The time resolution $\sigma^2 = \int  t^2 \ov \rho(t) dt - (\int  t \ov \rho(t) dt)^2$ has then two components, a contribution $\sigma_{av}$ from avalanche fluctuations and a contribution  $\sigma_{pos}$ from the varying position of the primary e-h pair in the gain layer
\bea
   (\gamma v^*)^2 \sigma_{av}^2 & = &  \frac{\pi^2}{6}\\
   (\gamma v^*)^2 \sigma_{pos}^2 & = &  \int_0^d p_1(x_0) \left[ \ln h(x_0)\right]^2 dx_0 - \left(  \int_0^d p_1(x_0) \ln h(x_0) dx_0\right) ^2 \label{postion variation_variance} 
\eea
 In the case of a very small photon absorption length i.e.~$l_a \ll d$ the primary e-h pair will always be created at the very edge of the sensor which is equal to the situation of $x_0=0$ and the time resolution is given by $\sigma_{av}$. In the other extreme of $l_a \gg d$ there will be a uniform distribution for the position of the photon interaction in the gain layer and the contribution to the time resolution is given in Fig.~\ref{inverse}b. Whether the avalanche fluctuations or the position fluctuations dominate depends on the gain layer thickness and the electric field. We can conclude that the time resolution is well approximated by $\sigma_1 \approx c_0/\gamma v^*$, where the main dependence is given by the variation of $\gamma$ with sensor thickness and electric field (Fig.~\ref{inverse}a) while $c_0 \approx 1{-}3$ and $v^*$ is saturated at $\approx 0.1\,\mu$m/ps.  
 \\ \\
 The term $\sigma_{pos}$ can also be derived directly from the average avalanche growth. Neglecting avalanche fluctuations the primary e-h pair will simply trigger an average avalanche according to Eq.~\ref{tot_number_dominant_term} 
 \beq \label{nofluct_approx}
          N(t) = B_e \frac{u_e (x_0) + u_h(x_0)}{P(x_0)}  e^{\gamma v^* t} 
 \eeq
where we have divided by the efficiency to account for the avalanches that do not cross the threshold. Applying a threshold $N_{thr}$ to this signal gives a threshold crossing time of  
 \beq \label{nofluct_approx}
          t(x_0)  = \frac{1}{\gamma v^*} \left[\ln N_{thr} - \ln h(x_0) \right]
 \eeq
and the variance $\ov{t^2} - \ov t^2$ is the one from  Eq.~\ref{postion variation_variance}.

\newpage


\section{Charged particle detection with SPADs \label{MIP_detection} }

\begin{figure}[h]
\begin{center}
     \epsfig{file=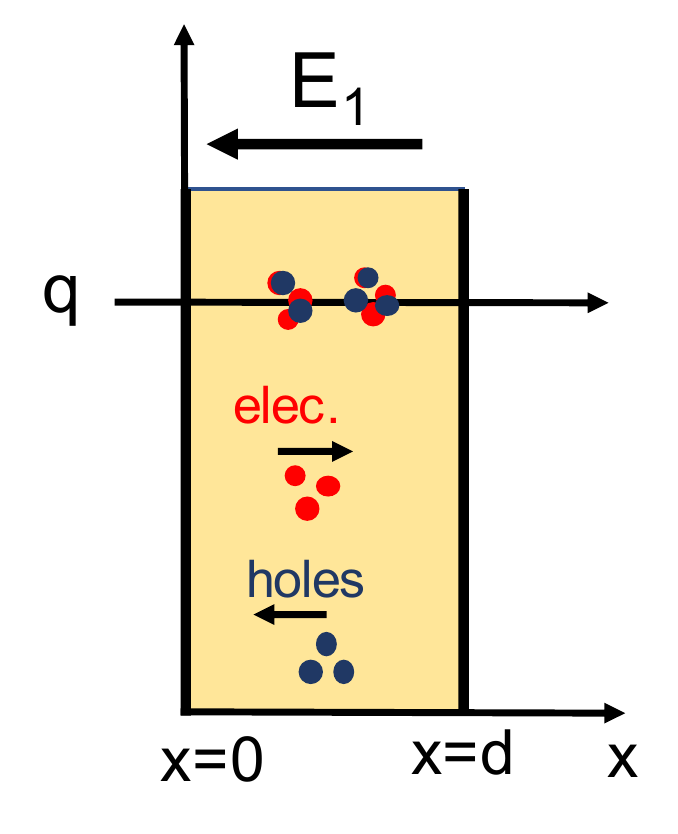, width=3cm}
\caption{A charged particle is leaving clusters of e-h pairs in the gain layer. }
\label{mip_detection}
\end{center}
\end{figure}
\begin{figure}[h]
\begin{center}
a)
     \epsfig{file=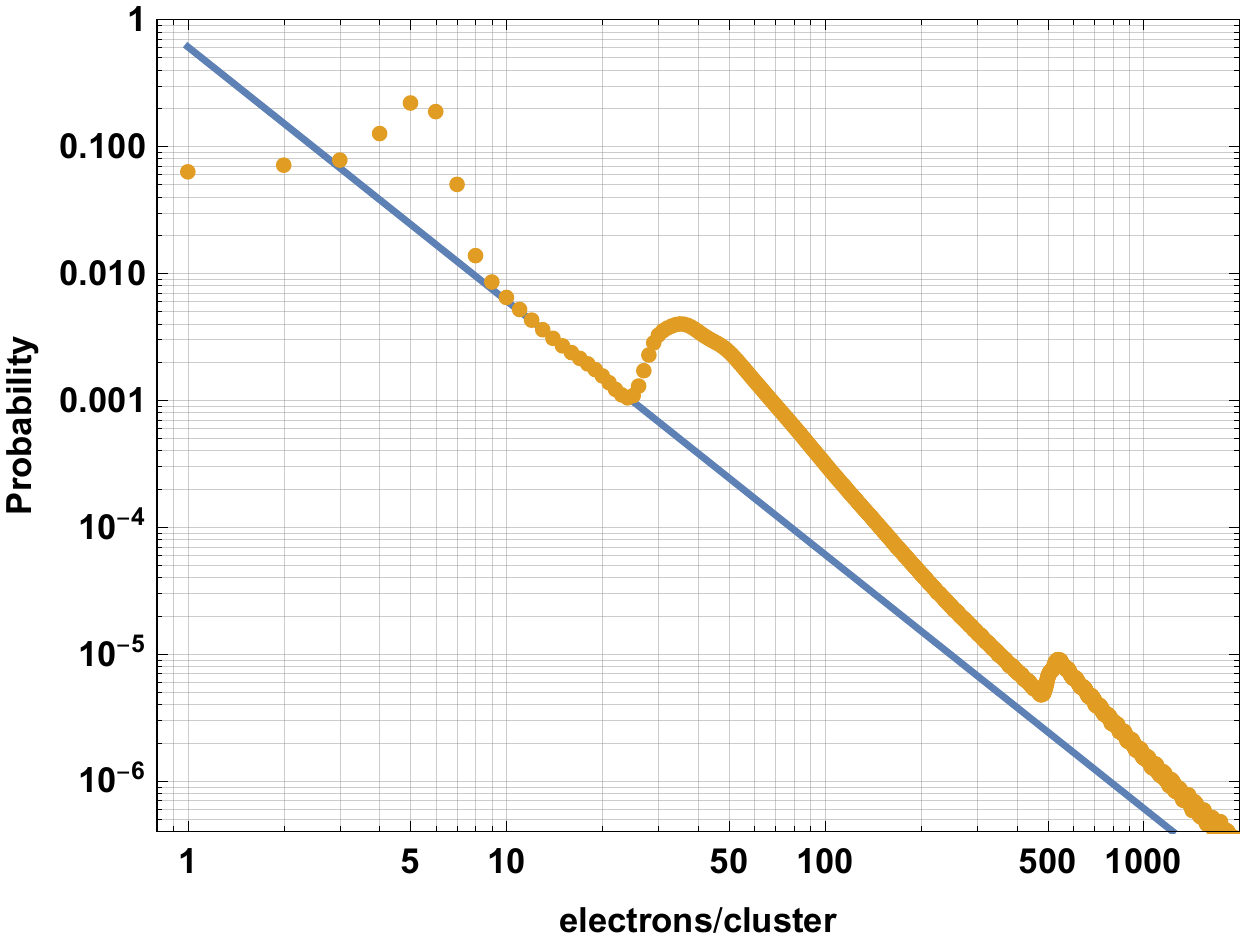, width=5cm}
     b)
      \epsfig{file=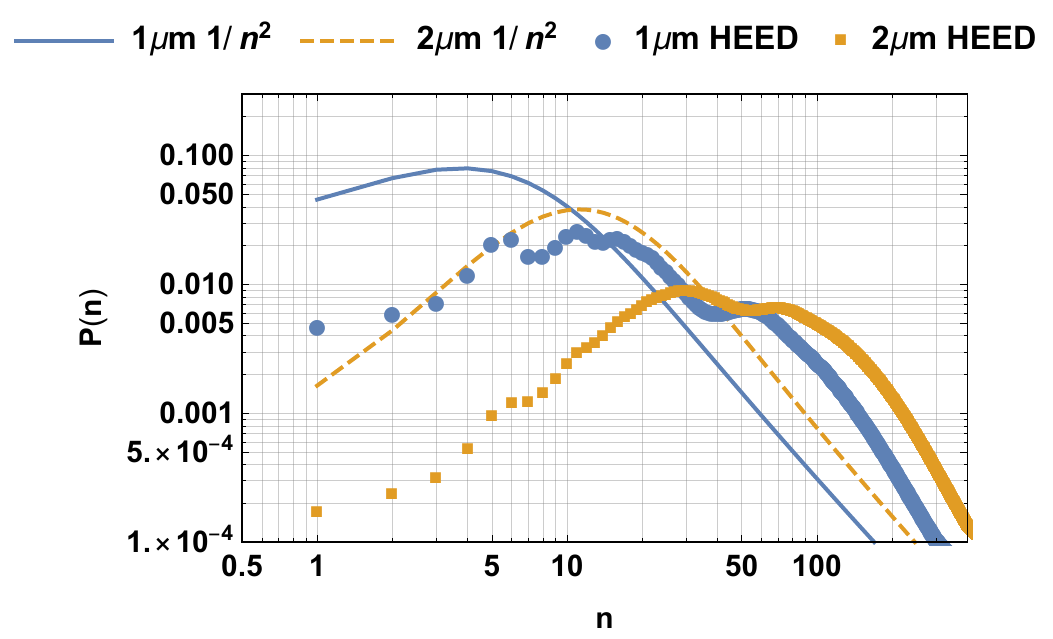, width=7cm}
\caption{a) Cluster size distribution i.e.~the probability for a single cluster to contain $n$ electron-hole pairs as calculated with HEED \cite{heed}. b) Probability $p(n, d)$ to find $n$ e-h pairs inside a silicon layer of thickness $d$. The solid lines refer to an $1/n^2$ distribution in both plots.   }
\label{mip_cluster_size}
\end{center}
\end{figure}

SPADs can also be used for detection of charged particles. Charged particles interacting with silicon produce clusters of e-h pairs along their track, with an average distance of $\lambda \approx 0.21\,\mu$m for MIPs.  A SPAD sensor of 1\,$\mu$m or 2\,$\mu$m thickness will therefore be highly efficient to charged particles and there is no need for a conversion layer. The probability $p_{clu}(n)$ for having $n>0$ e-h pairs in a single cluster is approximately given by a $1/n^2$ distribution, but there are significant deviations at small numbers of $n$ in silicon. Fig.~\ref{mip_cluster_size}a shows the cluster size distribution for silicon as calculated with HEED \cite{heed}. 
Assuming a sensor thickness $d$ and an average distance between clusters of $\lambda$, the average number of clusters in the sensor is $n_0 = d/\lambda$. Assuming the cluster size distribution $p_{clu}(n)$, the probability $p(n, d)$ to find $n$ e-h pairs in the sensor can then be calculated by using the Z-transform as \cite{RPC_werner}
\beq \label{pseudolandau}
P_{clu}  (z) =\sum_{n=1}^\infty \frac{p_{clu}(n)}{z^n} \qquad   G(z) = \frac{e^{d/\lambda P_{clu} (z)}-1}{e^{d/\lambda}-1}  \qquad p(n, d) = \frac{1}{n!} \left[ \frac{d^n G(1/z)}{d z^n}\right]_{z=0}
\eeq
It is shown in Fig.~\ref{mip_cluster_size}b.

\newpage

\subsection{Efficiency}

First we want to calculate the efficiency for a charged particle to cause breakdown in the gain layer. 
Since $P(x)$ from Eq.~\ref{breakdown_ehpair} is the probability for a single e-h pair at position $x$ to trigger a diverging avalanche, the probability $Q(x)$ that this single e-h pair does not cause breakdown is
\beq
     Q(x) = 1-P(x) = \frac{1}{1+\frac{p_0}{1-p_0}\exp \left[-\int_0^x(\alpha(x')-\beta(x'))dx' \right]}
\eeq
The probability $dq$ that there is no breakdown caused by the charged particle traversing the interval $[x, x+\Delta x]$ is given by a) the probability $1-\Delta x/\lambda$ that there is no interaction in $\Delta x$ and b) the probability that there is an interaction but it does not lead to breakdown, which we can write as 
\beq
      dq    =  \left(1-\frac{\Delta x}{\lambda} \right) +  \frac{\Delta x}{\lambda} \sum_{n=1}^\infty p_{clu}(n) Q^n(x) \\
\eeq
From this, the probability $q$ that the charged particle does not create a diverging avalanche in any of the slices $\Delta x$ is derived in Appendix D and the efficiency $p=1-q$ is given by
\beq \label{MIP_efficiency} 
     p = 1-\exp \left[  -\frac{1}{\lambda} \left(d - \sum_{n=1}^\infty p_{clu}(n) \int_0^d Q(x)^n dx \right) \right]
\eeq 
If $Q(x)=0$ i.e.~if an e-h pair deposited at $x$ will definitely cause breakdown, we have $p=1-e^{-d/\lambda}$, which is the correct probability that there is at least one interaction within $d$. The evaluation for constant $\alpha, \beta$ is given in Appendix E and shown in Fig.~\ref{eff2} together with the efficiency for a single electron starting at $x_0=0$ in the gain layer. Above the breakdown field the efficiency rises steeply to the maximum level $1-e^{-d/\lambda}$. For a gain layer thickness of $d>1\,\mu$m the efficiency is larger that 99\,\% above the breakdown limit.

\begin{figure}[h]
\begin{center}
     \epsfig{file=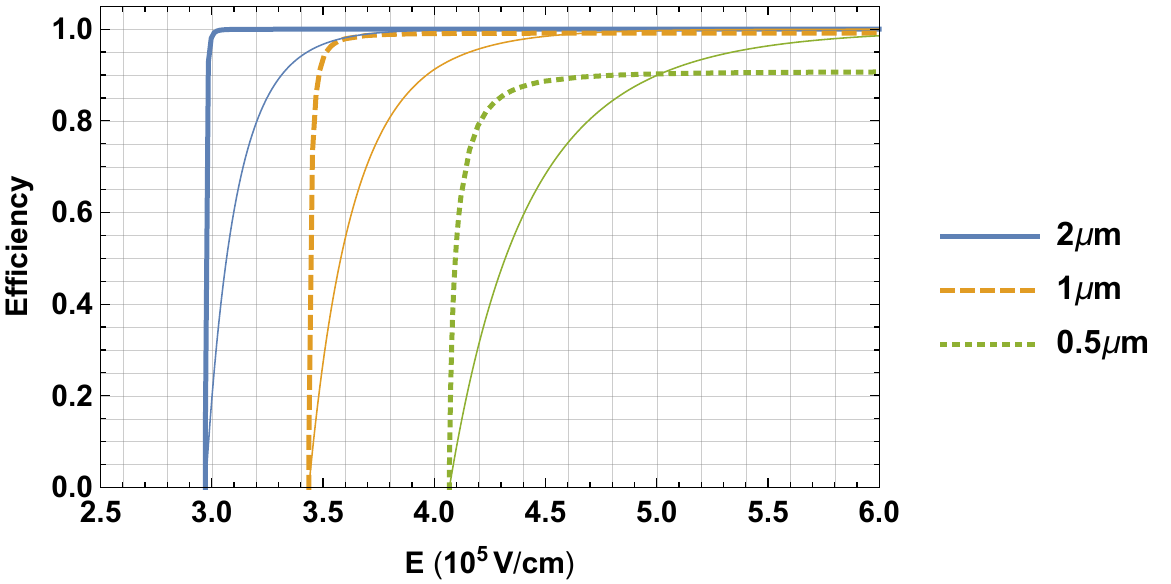, width=8cm}
\caption{Efficiency for a MIP to provoke breakdown. Beyond the breakdown field, the efficiency rises sharply from 0 to the level of $1-e^{-d/\lambda}$. The thin lines correspond to the efficiency for a single electron at $x_0=0$.}
\label{eff2}
\end{center}
\end{figure}


\newpage

\subsection{Time resolution}
\label{sec:timeres_bounded}

\begin{figure}[h]
\begin{center}
    a) \epsfig{file=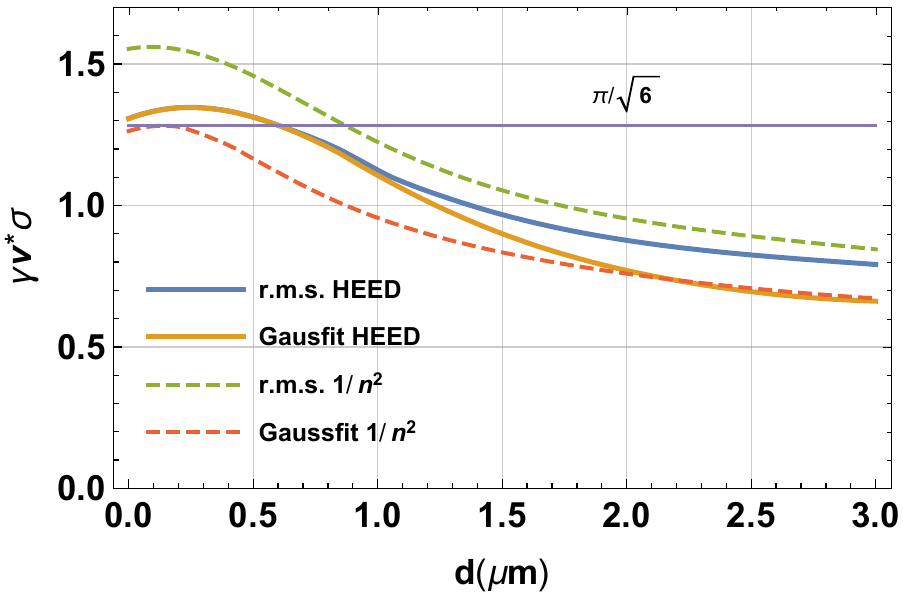, height=4cm}
    b)      \epsfig{file=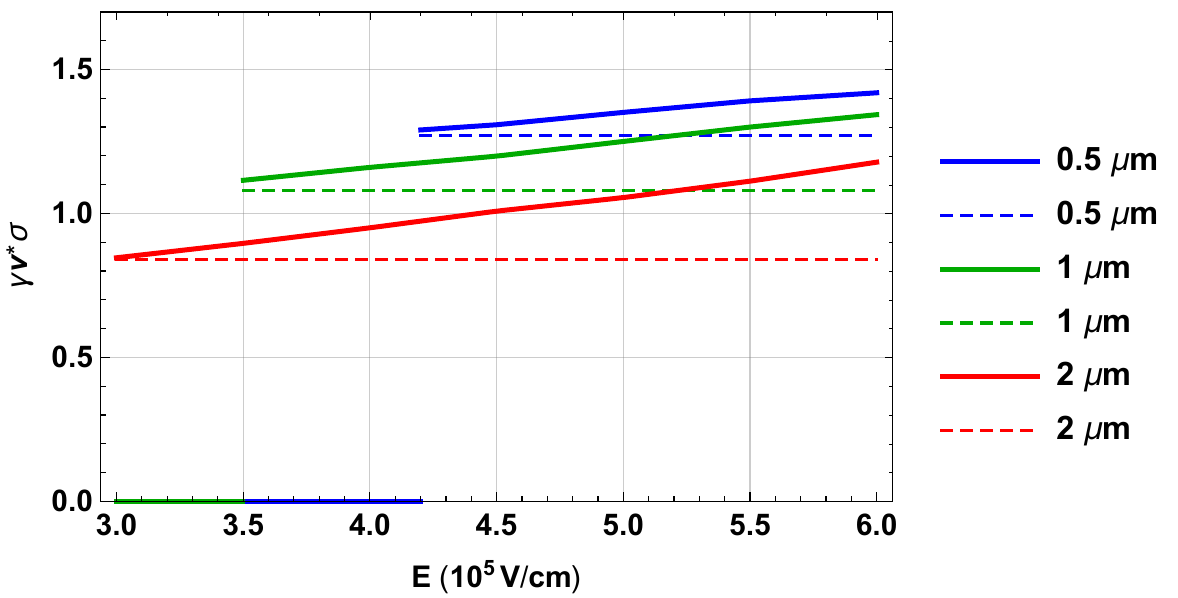, height=4cm}
\caption{ a) Time resolution for a MIP when the fluctuations of the cluster position can be neglected, for different values of the gain layer thickness $d$. The solid lines refer to $p_{clu}(n)$ for silicon from Fig.~\ref{mip_cluster_size}a, while the dashed lines assume $p_{clu}(n)\sim 1/n^2$. The time resolution is quantified by the standard deviation of the threshold crossing time (``r.m.s.'') as well as by the parameter $\sigma$ extracted from a Gaussian fit. b) Time resolution for a MIP for a gain layer of $0.5, 1, 2\,\mu$m thickness, taking the fluctuation of the cluster positions into account. The dashed lines refer to the numbers from a).}
\label{universal_reso_curve}
\end{center}
\end{figure}

We first consider the simpler case where the MIP produces a single cluster of $m$ e-h pairs at position $x_0$. Then, we have $A=m$, $n_e^0=n_h^0=m$ and the time response function is 
\beq
        \rho(t, m, x_0) \approx   \frac{ \gamma v^*}{(m-1)! }\ h(x_0)^m   \exp \left[ -m \gamma v^* t - h(x_0)e^{-\gamma v^*  t}\right]
\eeq
\beq
h(x_0) = \frac{1-[1-P(x_0)]^m}{ B_e [u_e(x_0)+u_h(x_0)]} \approx \frac{1}{ B_e [u_e(x_0)+u_h(x_0)]}
\label{h_x_defn}
\eeq
In case the cluster size varies according to $p_{clu}(m)$ and the probability to have the cluster at position $x_0$ varies according to $p_1(x_0)$, the time response function becomes
\beq
    \ov \rho(t) = \sum_{m=1}^\infty \int_0^d p_{clu}(m) p_1 (x_0) \rho( t, m, x_0) dx_0
\eeq
and the related time resolution has three contributions
\bea
      ( \gamma v^*)^2\sigma^2     
         & = &  \sum_{m=1}^\infty p_{clu}(m) \psi_1(m) \\
       &+& \sum_{m=1}^\infty p_{clu}(m) \psi_0(m)^2 - \left(  \sum_{m=1}^\infty p_{clu}(m)\psi_0(m) \right) ^2 \\
           & + &      \int_0^d p_1(x_0) [ \ln h(x_0) ]^2dx_0 - \left(   \int_0^d p_1(x_0) \ln h(x_0) dx_0   \right) ^2 \label{x0_dependence_timeres}
\eea
The first term represents the average of the avalanche fluctuations, where $\psi_1(m)$ is decreasing from $\psi_1(1)=\pi^2/6$ to zero with $\approx 1/m$ dependence. For the cluster size distribution in silicon from Fig.~\ref{mip_cluster_size}a this first term evaluates to $\approx 0.36$. 
\\
The second term represents the fact that an avalanche starting with $m$ e-h pairs will on average grow as $m e^{\gamma v^* t}$. The term evaluates to $\approx 1.39$ for the cluster size distribution in silicon, significantly larger than the contribution from the avalanche fluctuations. \\
The third term represents the dependence on the position of the primary cluster and for a uniform probability this term evaluates to the values already shown in Fig.~\ref{inverse}b.
%
%
%
%
%
%
\\
In conclusion we therefore observe that assuming a single e-h cluster at a random position in the gain layer, the avalanche fluctuations are negligible and only the average growth of the avalanche as well as the position dependence play a role. In a regime where the contribution from the position dependence is negligible ($<1$) only the fluctuation of the total charge is important, which results in a universal  dependence of the time resolution on the thickness of the gain layer 
\beq
      (\gamma^2 v^*)^2\sigma^2  =  \sum_{m=1}^\infty p(m, d) \psi_0(m)^2 - \left(  \sum_{m=1}^\infty p(m, d) \psi_0(m) \right) ^2  
\eeq
Here, $p(m,d)$ is the probability that the passing MIP produces $m$ e-h pairs in the gain layer of thickness $d$. The resulting time resolution is shown in Fig.~\ref{universal_reso_curve}a. Since $\gamma v^* \sigma$ is close to unity for typical dimensions of the gain layer, the time resolution of a SPAD for a MIP is essentially defined only by $\gamma$ and $v^*$ and the values are the ones given in Fig.~\ref{inverse}a. Since the cluster size distribution $p_{clu}(m)$ has a long tail towards large values of $m$, the same is true for the time response function. The standard deviation of the threshold crossing time is then generally not identical to the parameter $\sigma$ extracted from a Gaussian fit to the distribution of the threshold crossing time. Both measures are compared in Figure \ref{universal_reso_curve}a.
\newpage
Next, we consider the general case where the MIP produces a variable number of clusters, all of which fluctuate in size. Since avalanche fluctuations are negligible we perform this calculation by following the discussion around Eq.~\ref{nofluct_approx}. We divide the gain layer into $N+1$ slices of thickness $\Delta x= d/(N+1)$ and assume that a charged particle leaves $m_n$ primary e-h pairs in the $n^{th}$ slice. The efficiency for this case is very close to unity, and the additional small dependence on the cluster distribution is neglected here. Then, the total average number of charges in the avalanche becomes
\beq
N_{tot}(t) = \sum_{n=0}^N  m_n h \left( n\Delta x \right) e^{\gamma v^* t}
\eeq
with $h(x)$ taken from Eq.~\ref{h_x_defn}.
Applying a threshold $N_{thr}$ to this signal and shifting it by a constant offset of $\ln N_{thr}$ we find a threshold crossing time of 
\beq
         t(m_0,m_1,...,m_N) = -\frac{1}{\gamma v^*}\ln  \left[\sum_{n=0}^N  m_n h \left( n\Delta x \right) \right]
\eeq
The probability $p(m, \Delta x)$ to find $m$ e-h pairs in a slice $\Delta x$ is given by
\beq \label{slice_probability}
              p(m, \Delta x) =\left( 1 -\frac{\Delta x}{\lambda} \right) \delta_{m0} +\frac{\Delta x}{\lambda} p_{clu}(m)
\eeq
so the average threshold crossing time and the second moment are 
\bea \label{position_averages}
        \ov t  & = & - \lim_{N \rightarrow \infty}\frac{1}{\gamma v^*} \sum_{m_0=0}^\infty    \sum_{m_1=0}^\infty  ... \sum_{m_N=0}^\infty   p(m_0,\Delta x) p(m_1,\Delta x)... p(m_N,\Delta x) \ln  \left[\sum_{n=0}^N  m_n h \left( n\Delta x\right) \right] \\     
         \ov {t^2} & = &  \lim_{N \rightarrow \infty}\frac{1}{(\gamma v^*)^2} \sum_{m_0=0}^\infty    \sum_{m_1=0}^\infty  ... \sum_{m_N=0}^\infty   p(m_0,\Delta x) p(m_1,\Delta x)... p(m_N,\Delta x) \ln^2  \left[\sum_{n=0}^N  m_n h \left( n\Delta x \right) \right]
\eea
These relations are evaluated in Appendix F, giving a contribution to the time resolution of
\beq
     (\gamma v^*)^2 \sigma^2=  \left[ \int_0^\infty w_0(y) \ln^2  y dy - \left( \int_0^\infty w_0(y) \ln y dy \right)^2\right]
\eeq
with
\beq
     W_0(s) = \frac{ \exp \left[ \frac{d}{\lambda}\,\frac{1}{d} \int_0^d P_{clu} \left( s \frac{h(x)}{h(0)} \right)dx\right] - 1}{e^{d/\lambda}-1}
     \qquad
      \qquad  w_0(y) = L^{-1} [W_0(s)] 
\eeq
Here, $P_{clu}(s)$ is the Laplace transform of the cluster size distribution and the operator $L^{-1}$ denotes the inverse Laplace transform. The evaluation is shown in Fig.~\ref{universal_reso_curve}b. The dashed lines show the time resolution from Fig.~\ref{universal_reso_curve}a where the contribution from position fluctuations is neglected. We see that for fields around the breakdown limit the effect from the positions variations is small and it increases with increasing field. Overall, the time resolution for MIPs stays within $\sigma_t=(0.8{-}1.5)/\gamma v^*$ for the parameters investigated.



\section{Realistic field configuration \label{real_field_section}}

\begin{figure}[h]
\begin{center}
a)
\epsfig{file=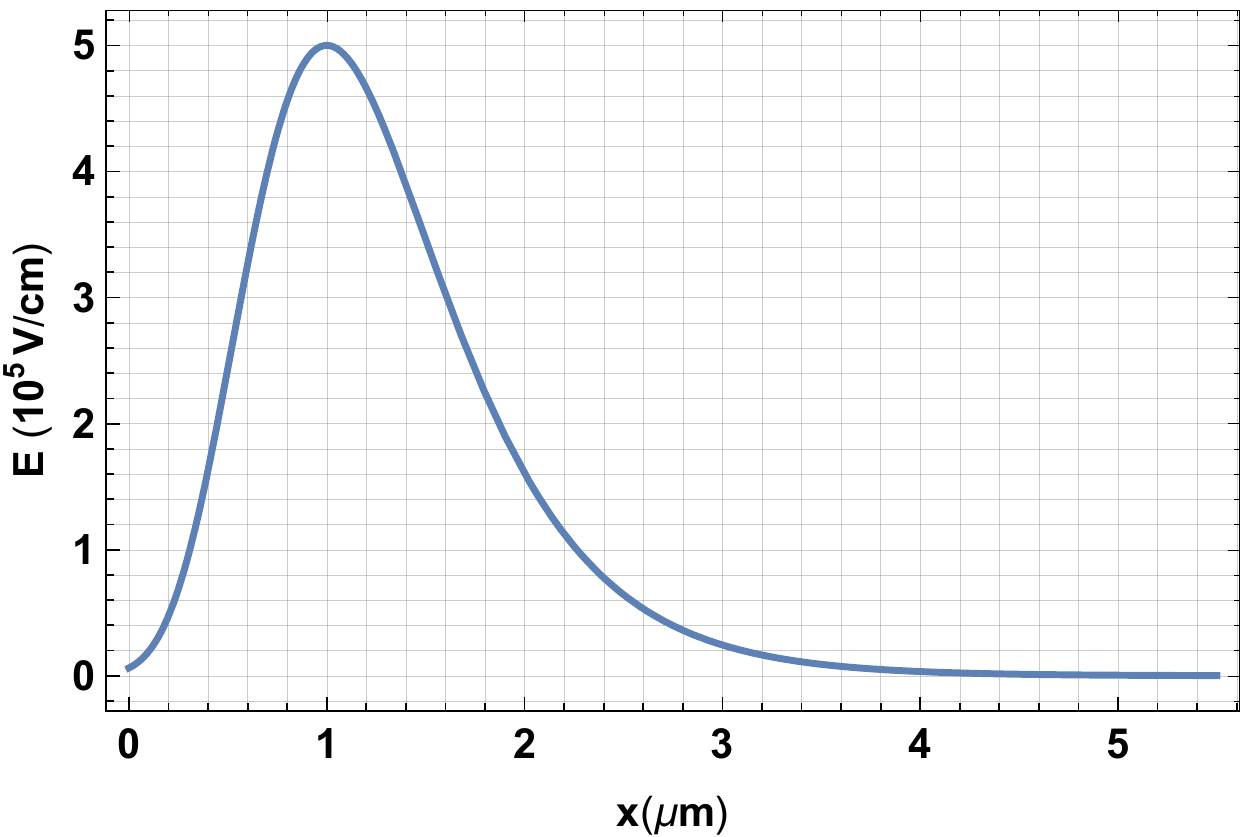, width=6cm}
b)
\epsfig{file=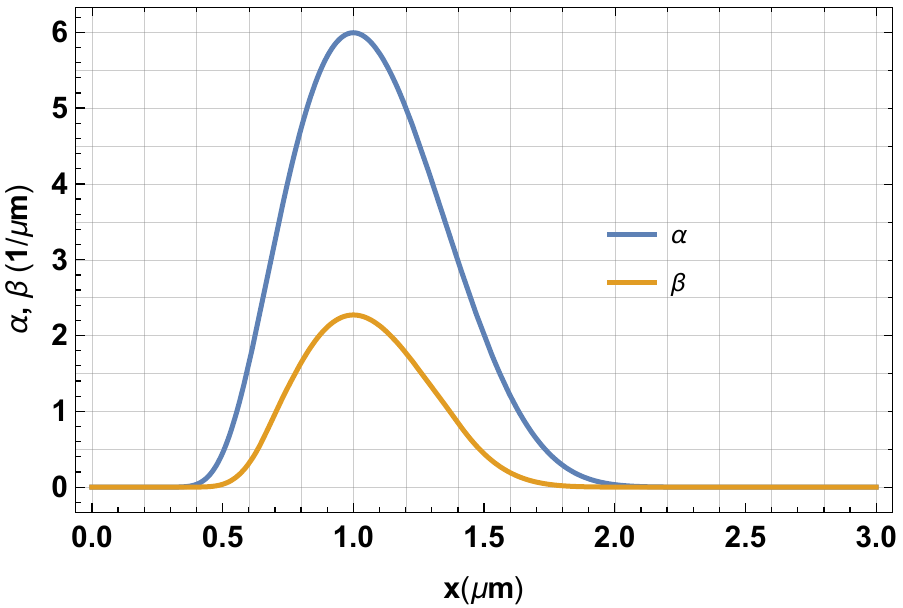, width=6cm}
\caption{a) Electric field in a realistic SPAD or SiPM. b) Impact ionization coefficients inside the sensor with parameters from Appendix A. }
\label{real_field}
\end{center}
\end{figure}

In this section we finally discuss a realistic field configuration of a SPAD and we apply the insights from all previous sections to assess its performance. Fig.~\ref{real_field}a shows an example for the electric field in a SPAD created by a highly doped p-n junction. The specific functional form is defined in Appendix A, Eq.~\ref{functional_field}. With the impact ionization and drift velocity parameters from Appendix A  we obtain $\alpha(x)$, $\beta(x)$, $v_e(x)$ and $v_h(x)$ as explicit functions throughout the sensor. The impact ionization coefficients $\alpha(x)$ and $\beta(x)$ are shown in Fig.~\ref{real_field}b. For the purposes of our discussion here, we define $x_1=0.4\,\mu$m and $x_2=1.9\,\mu$m as the boundaries of the gain layer, which thus has a thickness of $d=1.5\,\mu$m.

\begin{figure}[h!]
\begin{center}
a)
  \epsfig{file=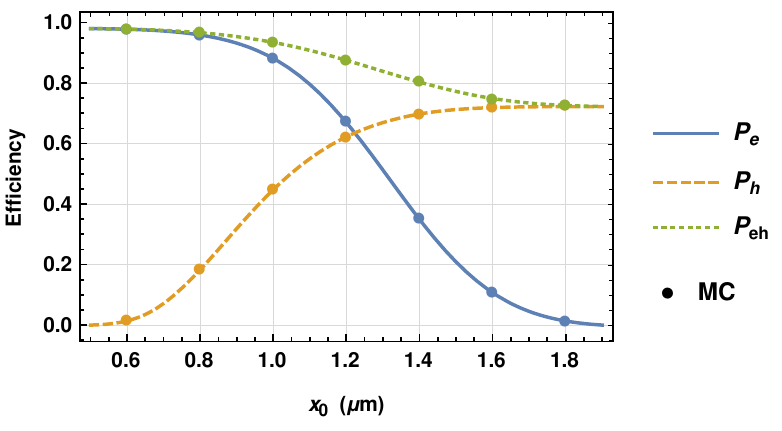, width =7cm}
  b)
  \epsfig{file=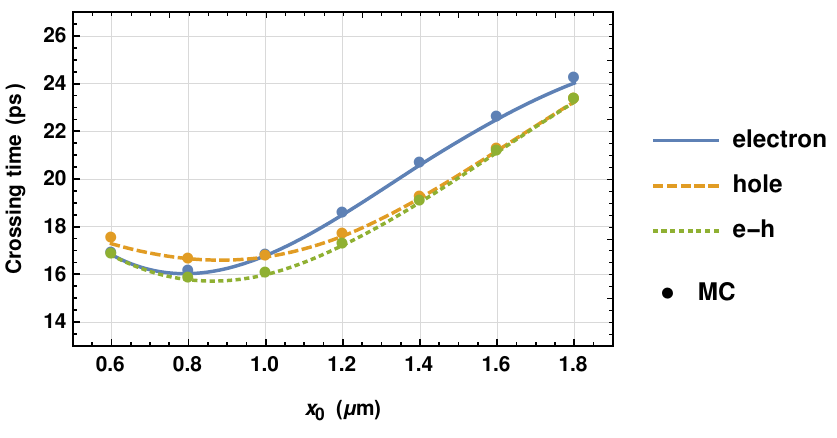, width=7.3cm}  
\caption{a) Breakdown probability for a primary electron ($P_e$), primary hole  ($P_h$) and primary e-h pair $(P_{eh})$ placed at position $x_0$ in the gain layer. b) Time at which the average total number of charges in the avalanche (proportional to the average signal) crosses a threshold of $10^4$ charges. The avalanche is initiated by a primary electron, a primary hole or a primary e-h pair placed at position $x_0$ in the gain layer. For both plots, the numerical solutions give rise to the lines, the markers correspond to the values from MC simulations. }
\label{real_peh}
\end{center}
\end{figure}

\paragraph{Efficiency}
For this geometry, the integral in Eq.~\ref{breakdown_condition} evaluates to 1.39 which is larger than unity and therefore guarantees that breakdown can take place. To find the efficiency of the sensor we solve Eqs.~\ref{turnon_equations} numerically, using as boundary conditions $P_h(x_1)=0$ and $P_e(x_2)=0$. The solution is shown in Fig.~\ref{real_peh}a together with the corresponding efficiencies obtained from a MC simulation of the avalanche development. The efficiency for a MIP passing this sensor can be calculated with Eq.~\ref{MIP_efficiency} using $Q(x) = 1-P_{eh}(x)$. With the cluster size distribution from Fig.~\ref{mip_cluster_size} and $\lambda = 0.21\,\mu$m, the efficiency evaluates to $p=1-8.5\times 10^{-4}$. A SPAD of this kind is a highly efficient detector for a MIP.
\paragraph{Average signal and contribution to time resolution}
To study the average growth of the avalanche, we solve Eqs.~\ref{average_equations} with $n_e(x_1) = 0$ and $n_h(x_2) = 0$ as boundary conditions. This yields the average charge densities in the gain layer, $n_e(x, t)$ and $n_h(x, t)$. The average total charge present in the gain layer can be obtained through a numerical integration of these densities. This quantity is proportional to the average signal produced by the avalanche. As shown in Section \ref{avalanches}, it grows exponentially as $e^{S t}$. The time constant $S$ can be directly extracted from the numerical solution and evaluates to $S = 0.48$\,ps.

\begin{figure}[h!]
\begin{center}
  \epsfig{file=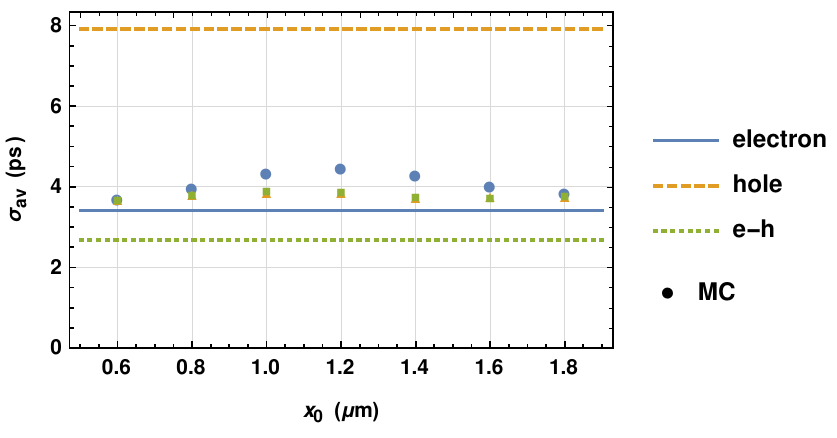, width=8cm}
  \caption{Time resolution $\sigma_{\mathrm{av}}$ due to avalanche fluctuations, for an avalanche initiated by a primary electron, a primary hole or a primary e-h pair. The primary charge is placed at position $x_0$ in the gain layer. The approximation from Eq.~\ref{timeres_bounded_eqn_approx} neglects the $x_0$-dependence and gives rise to the horizontal lines. The markers correspond to the values obtained from MC simulations.}
\label{real_time_resolution}
\end{center}
\end{figure}

In case the position $x_0$ of the primary charge fluctuates, it generates a contribution to the time resolution according to Eq.~\ref{x0_dependence_timeres}. The magnitude of this effect can be estimated from Fig.~\ref{real_peh}b, which shows the time at which the average signal crosses the applied threshold. If the position $x_0$ of the initial charge is uniformly distributed, the resulting contribution to the time resolution is 2.7/2.1/2.4\,ps for an initial electron, an initial hole, and an initial e-h pair.
\paragraph{Avalanche fluctuations and contribution to time resolution}
According to the discussion in Section \ref{sec:timeres_bounded}, we expect the contribution of the time resolution from fluctuations in the avalanche development, $\sigma_{\mathrm{av}}$, to be of the order of $1/S \approx 2$\,ps. A more precise estimate of the time resolution takes into account the primary charge initiating the avalanche. Following the discussion leading to Eq.~\ref{timeres_bounded_eqn_approx}, we approximate $\sigma_{\mathrm{av}}\approx\sqrt{\psi_1(A)} / S$. This formula neglects effects due to the finite size of the gain region and was originally derived for constant impact ionization coefficients, which enter into the computation of the parameter $A$. For position-dependent electric fields, the largest values of $\alpha$ and $\beta$ in the gain layer are relevant for the formation of the avalanche fluctuations and can be used to compute $A$. As the comparison with results from MC in Fig.~\ref{real_time_resolution} shows, this estimates the time resolution for the case of an initial electron to within at most 20\%. As expected, the approximation becomes worse if the initial charge includes holes, which have a low impact ionization coefficient. In this case, corrections due to the finite size of the gain layer become important. These can also be computed numerically, as shown in \cite{philipp_paper}, but the calculations are more involved.


\clearpage

\section{Conclusions}

We have performed a detailed study of the time resolution and efficiency of SPADs and SiPMs for the detection of photons and charged particles. Our discussions start from a series of differential equations, which cover the conversion and the drift of charges in the conversion layer as well as the formation of the avalanche in the gain layer. For arbitrary electric field profiles, the equations for the average avalanche development as well as the breakdown efficiency can be easily solved using numeric solvers. The calculation of the avalanche fluctuations and their impact on the time resolution is more involved in this case and a detailed discussion is given in \cite{philipp_paper}. We have provided analytic solutions for the case of constant electric fields.
\\ \\
For the detection of single photons, the contribution of the conversion layer of thickness $w$ to the time resolution for constant electric field is $\sigma=w/(v_e \sqrt{12})$. This is valid in case the photon absorption length is much larger than the conversion layer thickness, which simply corresponds to a uniform distribution of the photon conversion point inside the layer. 
\\ \\
The contribution of the gain layer to the time resolution has the general form
\beq
        \sigma = \frac{c_0}{\gamma v^*} 
\eeq
with $c_0=0.8{-}2.5$ for silicon and a gain layer thickness of $0.5{-}2\mu$m. This relation holds for single photon detection and MIP detection. It also extends to realistic non-uniform electric fields. Both contributions from avalanche fluctuations as well as the variation of the photon or MIP conversion point in the gain layer are captured. The constant $\gamma v^*$ determines the average growth of the avalanche according to $N(t) \propto e^{\gamma v^* t}$, with $v^* \approx 0.1\mu$m/ps and $\gamma$ saturating at $\gamma_{max} \approx \alpha+\beta$ at high fields.  
It should be possible in practice to limit this contribution to the level of a few picoseconds at high fields.
\\ \\
The efficiency of a SPAD or SiPM for photons has many contributions, including the photon conversion probability, the geometry and fill factor of the sensor as well as the breakdown probability in the gain layer. The contribution from the breakdown probability can be easily calculated by numerically solving the related equations.  For SPADs with a conversion layer or for photons with absorption length $<1\,\mu$m that are absorbed close to the edge of the gain layer this efficiency quickly approaches values close to 100\% when biasing the sensor beyond the breakdown field. 
\\ \\
SPADs or SiPMs with a gain layer of $1{-}2\,\mu$m thickness should be highly efficient for MIP detection. A dedicated conversion layer is not necessary.
\\ \\
This report discussed 'one-dimensional' sensors. For realistic implementations of SPAD pixel sensors and SiPMs the boundaries of the pixels together with all the elements for limitation of optical crosstalk make up complex three dimensional electric fields. To study these sensors, the 3D field map together with a full MC simulation with programs like Garfield++ \cite{garfieldpp} has to be used and our results can serve as benchmarks for these simulations.

\clearpage
\newpage


\section{Appendix A}

\subsection{Velocity of electrons and holes in silicon}

The velocity of electrons and holes in silicon is shown is parametrized by
\beq
     v_e(E) = 
     \frac{
         \mu_e\,E
     }{
         \left[
         1+\left(\frac{\mu_e\,E}{v^e_{sat}}\right)^{\beta_e}
         \right]^{1/\beta_e}    
     } 
     \qquad \qquad
     v_h(E) = 
     \frac{
         \mu_h\,E
     }{
         \left[
         1+\left(\frac{\mu_h\,E}{v^h_{sat}}\right)^{\beta_h}
         \right]^{1/\beta_h}    
     } 
\eeq
The parameters from \cite{canali} are $\mu_e=1417$\,cm$^2$/Vs, $\mu_h=471$\,cm$^2$/Vs, $\beta_e=1.109$, $\beta_h=1.213$ and $v^e_{sat}=1.07\times10^7$\,cm/s and $v^h_{sat}=0.837\times10^7$\,cm/s at 300\,K . 

\subsection{Impact ionization coefficients for electrons and holes in silicon }

The impact ionization parameters  $\alpha$ and $\beta$ as reported in \cite{overstraeten_man} are given by 
\beq
   \alpha(E) = \alpha_\infty e^{-a/E} \qquad  \beta(E) = \beta_\infty e^{-b/E}
\eeq
with
\bea
      \alpha_\infty = 7.030 \times 10^5\,cm^{-1} && a = 1.231\times 10^6 \, V/cm \qquad 1.75\times 10^5 \le E \le 6.0\times 10^5 \,V/cm \\
      \beta_\infty = 1.582\times10^6\,cm^{-1} && b=2.036\times10^6 \, V/cm  \qquad 1.75\times 10^5 \le E \le 4.0\times 10^5 \,V/cm \\
       \beta_\infty = 6.710\times10^5\,cm^{-1} && b=1.693\times10^6 \, V/cm  \qquad 4.00\times 10^5 \le E \le 6.0\times 10^5 \,V/cm 
\eea

\subsection{Functional form of electric field}

For a realistic SPAD we assume the following electric field:
\beq \label{functional_field} 
   E(x) = E_0 \,\exp \left[ 1-(x-\mu)/\sigma - e^{-(x-\mu)/\sigma}  \right]
   \qquad
   E_0 = 5 \times 10^5\,\mbox{V/cm} 
   \quad
   \mu = 1\,\mu\mbox{m}
   \quad 
   \sigma = 0.5\,\mu\mbox{m}
\eeq


\section{Appendix B}

For $S=0$ Eq.~\ref{gamma_value_equation} reads as 
\bea
        [v_e(x) f(x)]'  & = &  \alpha (x) v_e(x)  f(x) + \beta(x) v_h(x) g(x) \label{breakdown1}  \\
        -[v_h(x) g(x)]' & = &   \alpha (x) v_e(x)  f(x) + \beta(x) v_h(x) g(x)  \label{breakdown2} 
\eea
with boundary conditions $ f(0) = 0$ and $g(d) =0$. Subtracting the two equations gives
\beq \label{expression}
    [v_e(x) f(x)]'+ [v_h(x) g(x)]' = 0 \qquad \rightarrow \qquad v_e(x) f(x)= -v_h(x) g(x) + c_1
\eeq
Inserting this expression into Eq.~\ref{breakdown1} we have  
\beq
    [v_h(x) g(x)]' -v_h(x) g(x)[\alpha(x)-\beta(x)] = -c_1\alpha(x)
\eeq
with the general solution
\beq
    v_h(x) g(x) = \frac{ c_2-c_1\int_0^x \alpha(x') \exp [ -\int_0^{x'} (\alpha(x'')-\beta(x'')) dx''] dx' }
    {\exp[-\int_0^{x} (\alpha(x')-\beta(x')) dx']}
\eeq
The condition $f(0)=0$ implies $c_1=c_2$ and $g(d)=0$ then implies
\beq \label{condition1} 
     \int_0^d \alpha(x) \exp \left[ -\int_0^{x} (\alpha(x')-\beta(x')) dx' \right] dx = 1
\eeq
This is the general breakdown condition which is independent of the electron and hole velocities. Expressing $v_h g$ from Eq.~\ref{expression} and inserting this expression into Eq.~\ref{breakdown2} we have
\beq
   v_h(x) g(x)=  -v_e(x) f(x) + c_1  \qquad [v_e(x) f(x)]'-v_e(x) f(x)[\alpha(x)-\beta(x)] = c_1\beta(x)
\eeq
and therefore 
\beq
v_e(x) f(x) =\frac{ c_1\int_0^x \beta(x') \exp [ -\int_0^{x'} (\alpha(x'')-\beta(x'')) dx''] dx' +C}
    {\exp[-\int_0^{x} (\alpha(x')-\beta(x')) dx']}
\eeq
The condition $f(0)=0$ gives $C=0$ and $g(d)=0$ gives
\beq
       \int_0^d\beta(x) \exp \left[ -\int_0^{x} (\alpha(x')-\beta(x')) dx' \right] dx =  \exp \left[ -\int_0^{d} (\alpha(x')-\beta(x')) dx' \right]
\eeq
which is equal to
\beq  \label{condition2} 
      \int_0^d\beta(x) \exp \left[ \int_x^{d} (\alpha(x')-\beta(x')) dx' \right] dx =1
\eeq
As shown in \cite{stillman_wolfe}  Eq.~\ref{condition1} and \ref{condition2} are identical because
\beq
        \int_0^d(\alpha(x)-\beta(x)) \exp \left[ -\int_0^{x} (\alpha(x')-\beta(x')) dx' \right] dx = 1 -  \exp \left[ -\int_0^{d} (\alpha(x')-\beta(x')) dx' \right] 
\eeq


\section{Appendix C}

The relation that determines the number of electrons $n$ and the number of holes $m$ at a given time when starting with a single electron is defined in Eq.~\ref{penmt} as
\bea
        p_e(n, m, t+dt) & = &  (1-\alpha v_e dt)  p_e(n, m, t) \\
        & + & \alpha v_e dt \sum_{i=1}^n \sum_{j=1}^i \sum_{r=1}^m \sum_{s=1}^r p_e(n-i-j,m-r-s,  t)p_e(i,r,t)p_h(j,s,t)  \nonumber
\eea
Establishing the corresponding equation for $p_h(m, n, t)$ and expanding for small $dx$ we have
\bea
       \frac{1}{v_e} \frac{ d p_e(n, m, t)}{d t} & = &  -\alpha p_e(n, m, t) \label{pe_eqn_appendix}\\
        & + & \alpha \sum_{i=1}^n \sum_{j=1}^i \sum_{r=1}^m \sum_{s=1}^r p_e(n-i-j, m-r-s, t)p_e(i, r, t)p_h(j, s, t)  \nonumber
\eea
\bea
          \frac{1}{v_h} \frac{ d p_h(n, m, t)}{d t} & = &  -\beta  p_h(n, m, t) \label{ph_eqn_appendix}\\
        & + & \beta \sum_{i=1}^n \sum_{j=1}^i \sum_{r=1}^m \sum_{s=1}^r p_h(n-i-j,m-r-s, t)p_h(i, r, t)p_e(j, s, t)  \nonumber
\eea
The Z-transform of these equations is
\beq \label{eq_electrons_no_boundary} 
       \frac{1}{v_e} \frac{ \de P_e(z_1, z_2, t)}{\de t}  =   -\alpha P_e(z_1, z_2, t) [ 1- 
         P_e(z_1, z_2, t) P_h(z_1, z_2,  t)]
\eeq
\beq \label{eq_holes_no_boundary} 
            \frac{1}{v_h} \frac{ \de P_h(z_1, z_2, t)}{\de t}  =   -\beta  P_h(z_1, z_2, t) [ 1-
           P_h(z_1, z_2, t) P_e(z_1, z_2, t)  ]
\eeq
The equations have a structure similar to \ref{turnon_equations} and we therefore form the function
\beq
   P(z_1, z_2, t) = 1-P_e(z_1, z_2, t) P_h(z_1, z_2, t) 
\eeq
Differentiating this equation and using the above relations gives
\beq
    \frac{\de P}{\de t} = (\alpha v_e+\beta v_h) (1-P) P
\eeq
with the solution
\beq
    P(t) = \frac{e^{(\alpha v_e+\beta v_h)t}}{e^{(\alpha v_e+\beta v_h)t}+c_1}
\eeq
The initial conditions that there is one electron at $t=0$ for $p_e$ and one hole for $p_h$ reads as $p_e(n,m,t=0)=\delta_{n,1}\delta_{m,0}$ and $p_h(n,m,t=0)=\delta_{n,0}\delta_{m,1}$ and we have therefore 
\beq
  P_e(z_1, z_2, t=0) = \frac{1}{z_1} \qquad   P_h(z_1, z_2, t=0) = \frac{1}{z_2}  \qquad \rightarrow \qquad P(z_1, z_2, t=0) = 1-\frac{1}{z_1z_2}  
\eeq
and
\beq
    P(z_1, z_2, t) =   \frac{e^{(\alpha v_e+\beta v_h )t}(z_1z_2-1)}{1+e^{(\alpha v_e+\beta v_h)t}(z_1z_2-1)}
\eeq
We can now write Eqs.~\ref{eq_electrons_no_boundary} and \ref{eq_holes_no_boundary} as
\beq
      \frac{ d \ln P_e}{dt} = -\alpha v_e P  \qquad     \frac{ d \ln P_h}{dt} = -\beta v_h P
\eeq
Integrating the equations with the above initial conditions we finally have
\bea
    P_e(z_1, z_2,  t)  &= & \frac{1}{z_1} \left[ \frac{z_1z_2}{1+e^{(\alpha v_e+\beta v_h)t} (z_1z_2-1) }\right]^{\frac{\alpha v_e}{\alpha v_e+\beta v_h}}
 \\
    P_h(z_1, z_2,  t)  & = &  \frac{1}{z_2} \left[ \frac{z_1z_2}{1+e^{(\alpha v_e +\beta v_h)t} (z_1z_2-1) }\right]^{\frac{\beta v_h}{\alpha v_e+\beta v_h}}
\eea
In case the avalanche is initiated by $n_e^0$ electrons and $n_h^0$ holes at time $t=0$, we are interested in the probability $p_0(n,m,t)$ to find $n$ electrons and $m$ holes at time $t$. In terms of $p_e(n,m,t)$ and $p_h(n,m,t)$, it is expressed as an iterated convolution in analogy to the right-hand sides of Eqs.~\ref{pe_eqn_appendix} and \ref{ph_eqn_appendix}. In the $z$-domain, $P_0(z_1, z_2, t)$ reads
\bea
          P_0(z_1, z_2, t) & =  &  P_e(z_1, z_2,  t) ^{n_e^0}  P_h(z_1, z_2,  t) ^{n_h^0} \\
          & = & 
            \frac{1}{z_1^{n_e^0}}   \frac{1}{z_2^{n_h^0}} 
            \left[ \frac{z_1z_2}{1+e^{(\alpha v_e +\beta v_h)t} (z_1z_2-1) }\right]^{\frac{n_e^0 \alpha v_e + n_h^0 \beta v_h}{\alpha v_e+\beta v_h}}
\eea
The inverse Z-Transform of this expression also gives access to $p(n, t)$ shown in Eq.~\ref{p_additional_eh}, which is defined as the probability to find $n$ e-h pairs that are created in addition to the initial $n_e^0$ electrons and $n_h^0$ holes.
\beq
     p(n, t) = \frac{\Gamma(A+n)}{\Gamma (A) \Gamma (1+n) } \left(  \frac{1}{\nu (t)} \right)^A \left( 1 - \frac{1}{\nu (t)}  \right)^{n} \qquad
      \sum_{n=0}^\infty p(n,t)=1
\eeq
\beq
    A=\frac{n_e^0\alpha v_e+ n_h^0\beta v_h}{\alpha v_e+\beta v_h}  \qquad 
    \nu (t) = e^{(\alpha v_e+\beta v_h)t}  
\eeq


\section{Appendix D}

We divide the sensor into $N+1$ slices of thickness $\Delta x=d/(N+1)$. The probability that there is no breakdown caused by the particle traversing the slice $[x, x+\Delta x]$ is given by the probability $1-\Delta x/\lambda$ that there is no interaction in $\Delta x$ and the probability that there is an interaction but it does not lead to breakdown.
\bea
      dq   & = & \left(1-\frac{\Delta x}{\lambda} \right) +  \frac{\Delta x}{\lambda} \sum_{n=1}^\infty p_{clu}(n) Q^n(x) \\
                            & = &  1 - \frac{\Delta x}{\lambda} \left( 1 - \sum_{n=1}^\infty p_{clu}(n) Q^n(x) \right)  \\
                            & := & 1 - \frac{\Delta x}{\lambda} f(x) 
\eea
The probability $q$ that there is no breakdown in any of the slices of $\Delta x$ throughout the sensor is then
\bea
         q & = & \left[ 1 - \frac{\Delta x}{\lambda} f(0) \right] 
         \left[ 1 - \frac{\Delta x}{\lambda} f(\Delta x) \right]  
         \left[ 1 - \frac{\Delta x}{\lambda} f(2 \Delta x) \right] ...
          \left[ 1 - \frac{\Delta x}{\lambda} f(N \Delta x) \right]  \\
           \ln q & = & \sum_{n=0}^N \ln   \left[ 1 - \frac{d}{N \lambda} f(n d/N) \right] \\
             & \approx & \sum_{n=0}^N - \frac{d}{N \lambda} f(n d/N) \\
              & \approx & - \frac{1}{\lambda}  \int_0^\infty  f(x)dx \\
              & = & -\frac{1}{\lambda} \left(d - \sum_{n=1}^\infty p_{clu}(n) \int_0^d Q(x)^n dx \right)
\eea
and the probability $p=1-q$ that the sensor is efficient is therefore 
\beq
     p = 1-\exp \left[  -\frac{1}{\lambda} \left(d - \sum_{n=1}^\infty p_{clu}(n) \int_0^d Q(x)^n dx \right) \right]
\eeq


\section{Appendix E}

\beq
         Q(x) = 1-P(x) = \frac{1}{1+\frac{p_0}{1-p_0}\exp \left[-(\alpha -\beta)x \right]}
\eeq
\beq
     \int_0^d Q(x)^n dx =  \frac{1}{\alpha-\beta} \left[ H \left(n,1+\frac{p_0\exp [-(\alpha-\beta)d] }{1-p_0} \right)    - 
     H \left(n, 1+\frac{p_0}{1-p_0}  \right) \right]
\eeq
\beq
     H(n, y) = \int \frac{dy}{y^n(1-y)} = \ln \frac{y}{1-y} - \sum_{m=1}^{n-1} \frac{1}{m y^m} 
\eeq


\section{Appendix F}

We assume the probability $p_{clu}(n)$ to be continuous in $n$ (which we can imagine by expressing it as a sum of delta functions centered at integer values of $n$) and write the expression in Eq.~\ref{slice_probability} as 
\beq \label{slice_probability2}
              p(n, \Delta x) =\left( 1 -\frac{\Delta x}{\lambda} \right) \delta(n) +\frac{\Delta x}{\lambda} p_{clu}(n)
\eeq
We can then replace the sums in Eq.~\ref{position_averages} by integrals and have
\beq
        \ov t (N)  =  -\frac{1}{\gamma v^*} \int dm_0 \int dm_1 ... \int d m_N    p(m_0,\Delta x) p(m_1,\Delta x)... p(m_N,\Delta x) \ln  \left[\sum_{n=0}^N  m_n h \left( n\Delta x\right) \right]  
\eeq
We change variables according to 
\beq
      m = \frac{1}{h_0}\sum_{n=0}^N  m_n h _n \quad \rightarrow \quad m_0 = m - \frac{1}{h_0}\sum_{n=1}^N  m_n h_n 
\eeq
where we have written $h_n=h(n\Delta x)$, which gives 
\bea
   \ov t (N)  & = &   -\frac{1}{\gamma v^*} \int dm \left[\int dm_1 ... \int d m_N    p \left( m-\frac{1}{h_0}\sum_{n=1}^N  m_n h_n, \Delta x \right) p(m_1, \Delta x)... p(m_N, \Delta x) \right]\ln (h_0 m)  \no \\
   & = &    -\frac{1}{\gamma v^*} \int w(m) \ln (h_0 m)   \,  dm
\eea
with
\beq
      w(m) = \int dm_1\int dm_2 ... \int d m_N     p \left( m-\frac{1}{h_0}\sum_{n=1}^N  m_n h_n, \Delta x \right) p(m_1, \Delta x)... p(m_N, \Delta x)
\eeq
The Laplace transform of this expression is
\beq
     W(s) = P(s, \Delta x )P\left(  \frac{h_1}{h_0} s, \Delta x\right) P\left( \frac{h_2}{h_0} s, \Delta x\right) ... P\left(  \frac{h_N}{h_0} s, \Delta x\right)  = 
    \exp \left[ \sum_{n=0}^N \ln  P\left(  \frac{h_n}{h_0} s, \Delta x\right) \right]
\eeq
With $P(s, \Delta x)$ being the Laplace transform of Eq.~\ref{slice_probability2}
\beq
       P(s, \Delta x) = 1+\frac{\Delta x }{\lambda} (P_{clu}(s )-1) 
\eeq
we have
\bea
    W(s)  & = &    \exp \left( \sum_{n=1}^N \ln \left[ 1+\frac{\Delta x}{ \lambda } (P_{clu} \left( s \frac{h_n}{h_0}\right)-1) \right] \right)  \\
    & \approx & \exp \left( \sum_{n=1}^N \frac{\Delta x}{ \lambda} \left[ P_{clu} \left( s \frac{h_n}{h_0} \right)-1\right] \right) \\
      & = &  e^{-d/\lambda} \exp \left[ \frac{1}{\lambda} \int_0^d P_{clu} \left( s \frac{h(x)}{h(0)} \right)dx\right] 
\eea
Normalizing to the probability that there is at least one interaction inside the gain layer we have
\beq
     W_0(s) = \frac{ \exp \left[ \frac{d}{\lambda}\,\frac{1}{d} \int_0^d P_{clu} \left( s \frac{h(x)}{h(0)} \right)dx\right] - 1}{e^{d/\lambda}-1}
\eeq
The first and second moment of the threshold crossing time are therefore 
\beq
       w_0(y) = L^{-1} [W_0(s)] \qquad \ov t = -\frac{1}{\gamma v^*} \int w_0(y) \ln [h(0)y] dy \qquad  \ov {t^2}=  \frac{1}{(\gamma v^*)^2}\int w_0(y) \ln^2 [h(0)y]dy
\eeq
Comparing the expression to Eq.~\ref{pseudolandau} we see that the effect of the position dependence $h(x)$ is equivalent to a change of the cluster size distribution according to 
\beq
       \ov P_{clu}(s) =\frac{1}{d}\int_0^d P_{clu} \left( s \frac{h(x)}{h(0)} \right)dx
\eeq

\newpage


\end{document}